\documentclass[a4paper, fleqn,usenatbib]{mnras}
\usepackage{natbib}
\usepackage{pdflscape}
\usepackage{newtxtext,newtxmath}

\usepackage[T1]{fontenc}
\usepackage{ae,aecompl}
\usepackage{epsfig}

\newcommand{\msun}{M_{\odot}}	
\newcommand{\lambdar}{$\lambda(R_e)$}
\newcommand{\vsigma}{$V/\sigma$}
\newcommand{\vrotsigma}{$V_{ROT}/\sigma $}

\title[The role of environment in S0s formation.]{Formation of S0s in extreme environments III: the role of environment in the formation pathways.}

\author[L. Coccato et al.]{
Lodovico Coccato$^1$\thanks{E-mail: lcoccato@eso.org},
Amelia Fraser-McKelvie$^{2,3}$,
Yara L. Jaff\'e$^4$,
Evelyn J. Johnston$^5$,
Arianna Cortesi$^{6, 7}$, \and
Diego Pallero$^4$
\\
$^1$European Southern Observatory, Karl-Schwarzchild-str.,2, 85748 Garching b. M\:unchen, Germany.\\
$^2$International Centre for Radio Astronomy Research, The University of Western Australia, 35 Stirling Hwy, 6009 Crawley, WA, Australia.\\
$^3$ARC Centre of Excellence for All Sky Astrophysics in 3 Dimensions (ASTRO 3D).\\
$^4$Instituto de Fisica y Astronomia, Universidad de Valparaiso, Avda. Gran Breta\~na 1111 Valparaiso, Chile.\\
$^5$N\'ucleo de Astronom\'ia de la Facultad de Ingenier\'ia y Ciencias, Universidad Diego Portales, Av. Ejército Libertador 441, Santiago, Chile.\\
$^6$Observat\'orio do Valongo, Ladeira do Pedro Ant\^onio 43, CEP:20080-090, Rio de Janeiro, RJ, Brazil.\\
$^7$Centro Brasileiro de Pesquisas Físicas, Rua Dr. Xavier Sigaud 150, CEP 22290-180, Rio de Janeiro, RJ, Brazil.
}

\date{Accepted XXX. Received YYY; in original form ZZZ}

\pubyear{2022}

\begin{document}
\label{firstpage}
\pagerange{\pageref{firstpage}--\pageref{lastpage}}
\maketitle

\begin{abstract}
It is well established that there are at least two main channels to form lenticular (or S0) galaxies. The first, which we name "faded spiral" scenario, includes quenching events that led to consumption or removal of gas from a spiral progenitor. The second, which we call "merger" scenario, includes merger-like events and interactions between galaxies. Each scenario leaves characteristic signatures in the newly-formed lenticular galaxy. However, the conditions that trigger one mechanism over another are still unknown. This paper is the third of a series aimed at understanding the role of the environment in the formation of lenticular galaxies. In this study, we combine the kinematics, morphology, and properties of the stellar populations of 329 S0s from the SAMI and MaNGA surveys in order to highlight the role of the environment in the process. We divide the S0s into two classes (A and B) according to their global properties, that we can associate to the products of a faded spiral scenario (class A) or a merger scenario (class B).  We then study how the various classes are distributed within different environments. 
Our study reveals that the "faded spiral" pathway is the most efficient channel to produce S0s, and it becomes more efficient as the mass of the group or cluster or local density of galaxies increase. The merger pathway is also a viable channel, and its efficiency becomes higher with decreasing local density or environment mass.
\end{abstract}




\section{Introduction}
\label{sec:intro}
Lenticular galaxies (or S0s) are characterised by the presence of a spheroidal component and a stellar disk without spiral arms. Other features might be present (thick disk, shells, bars) depending on the evolutionary paths over time. From a formation point of view, it has been established that there are two main families of scenarios that can form S0s. The first includes processes such as mergers and interaction with other galaxies (e.g., \citealt{Tapia+17, Bekki+11}).  S0s formed in this way will show characteristics similar to those of elliptical galaxies: for example they would be, on average, more pressure supported than spirals of similar mass.  Not all mergers can form disk galaxies; in general major mergers tend to decrease the total angular momentum favoring the formation of spheroidal systems \citep{Navarro+94}. However, extremely massive disc galaxies in the nearby Universe can form through gas-rich minor mergers \citep{Jackson+22}. Moreover, under certain conditions, simulations shown that also major mergers can led to the formation of stable disk structures \citep{Governato+07, Chilingarian+10}. Moreover in certain cases, the S0-like merger remnants do not exhibit any morphological traces of the merger under typical observing conditions and at distances as nearby as 30 Mpc, 1–2 Gyr after the event \citep{Eliche-Moral+2018}.

The second family includes processes  such as quenching of star formation, starvation and ram pressure stripping, gas ejection by active nuclei, and, more generally, processes related to gas consumption and the suppression of star formation (e.g., \citealt{vanDenBergh09, Laurikainen+10, Rizzo+18}). These processes lead to the fading of the spiral arms in late type galaxies, so that only the stellar spheroid and the overall disk-like structures remain.

Because the "merger" and "faded spiral" scenarios comprise different physical processes, the resulting S0s are predicted to have different properties (see, for example, \citealt{Deeley+21, Yoon+21} and references therein).
Merger products tend to be more massive, older, less metal rich, more pressure supported, and have higher S\'ersic indices and lower ellipticities; sometimes they show the typical morphological features of past interactions (shells and tidal streams). 
Products from faded spiral arms tend to be less massive, younger, more metal rich, more rotationally supported and have lower S\'ersic indices.
Obviously, there is no well defined threshold in any physical property or their combination that uniquely determines the formation scenario, as a large number of  events might occur in a galaxy lifetime and a large number of factors, such as the presence of gas, AGN feedback, the density of intra-cluster medium and time itself, play fundamental roles in shaping the properties of each galaxy.

Among the various properties, stellar mass and kinematics seem to be those that can be more easily linked to a formation scenario. \citet{McDermid+15} studied the properties of Early Type Galaxies (ETGs, thus including both ellipticals and lenticulars) in the Atlas3D survey \citep{Cappellari+17}. They found that more massive ETGs are older, more metal rich, more $\alpha$-enhanced, and have a mild dependency with the environment, parametrised with the galaxy number density:  galaxies in the densest regions are overall older, more metal rich,  and more massive. Similarly, \citet{Fraser+18} studied the properties of the stellar populations of disks and bulges in $\sim 250$ lenticular galaxies selected from the MaNGA\footnote{Mapping Nearby Galaxies at Apache Point Observatory, \citealt{Bundy+15}.} survey. Their sample showed that galaxies with stellar mass above $10^{10}$ $\msun$ were preferentially old and metal rich, with bulges predominantly older than their discs, and most likely formed via merger events. On the other hand, galaxies with stellar mass below $10^{10}$ $\msun$ were preferentially more metal poor, with bulges displaying more recent star formation than their disks, and most likely formed via fading spiral scenario. Contrarily to \citet{McDermid+15}, they did not find significant dependency with local environmental properties; possibly because the MaNGA sample does not include over-dense cluster regions.
\citet{Rizzo+18} inferred the formation scenario of 10 S0s in the CALIFA galaxy survey from their angular momentum content, finding that mergers and interaction processes are not the dominant mechanism o S0 formation in the field and small groups. 

The stellar mass of a galaxy certainly influences the evolution of a system or its properties (e.g., its relation with metallicity, morphology, or location in a cluster environment). However, other elements play a role. Indeed, we can consider the stellar mass as a property of the final product, where the formation pathway has already been chosen by some other external factors or initial conditions. In this view, the question of what triggers a formation mechanism, or what the initial conditions that favour one scenario over another are, remains valid. Environment is the first candidate external factor that should be investigated to answer this question. Indeed, since the discovery of the morphology-density relation \citep{Dressler80} it has been shown that the environment is closely associated to galaxy formation mechanisms.


In our pilot study \citep{Coccato+20} we studied the properties of 21 lenticulars in different environments and found an indication that S0s in the cluster are more rotationally supported, suggesting that they are formed through processes that involve the rapid consumption or removal of gas (e.g. starvation, ram pressure stripping). In contrast, S0s in the field are more pressure supported, suggesting that minor mergers served mostly to shape their kinematic properties. 
\citet{Deeley+20} extended our study and confirmed that the two S0 formation pathways (merger and faded spiral) are active, with mergers dominating in isolated galaxies and small groups, and the faded spiral pathway being most prominent in large groups ($10^{13} \msun $  <M$_{\rm Halo Group}$<$10^{14} \msun $). 
In our second paper \citep{Johnston+21}, we concentrated of the formation time-scales of individual structural components (bulge, disk and lens). We found that, whereas bulge and disk are mostly coeval formed at high redshifts from the same material, lenses formed on independent time-scales, and possibly even from evolved bars.

It is fair to say that these studies do not consider the more complicated case in which
both scenarios occur for a single galaxy: a spiral can "fade" and passively evolve into an S0 and then experience a merger. Therefore,
when we refer to a formation scenario for an S0, we refer to the scenario that contributed to most of its present days properties.   
This does not, however, rule out the possibility that the galaxy experienced, at some point in its history, the physical processes of the other scenario, but implies that this only had a minor influence on the galaxy’s properties.

The purpose of this paper is to quantify the role of environment in the selection of a formation pathway for S0s. To this aim, we proceeded in the following way. First we identified two classes of S0s that share similar physical properties and then we linked each class with one formation scenario (merger or faded spiral). Finally, we measured the fraction of galaxies of each class in each environment (field, groups, cluster). If the environment plays a role in establishing the formation path for S0s, we expect that galaxies of a given class are distributed among the various environments differently from galaxies of the other class (depending, of course, on how strong the link between S0 class and formation scenario is).

The paper is organised as follows: Section \ref{sec:sample} presents our galaxy sample and the derived quantities. Section \ref{sec:analysis} describe our analysis, our definition of S0 classes, their relation with the formation scenarios, and how they are distributed with the environment. Comparison of our data with literature and discussion of potential biases are discussed in Section \ref{sec:comparison}. Finally, we summarise our work in Section \ref{sec:conclusion}.

\section{Physical properties of the sample galaxies}
\label{sec:sample}

In this Section we describe how the lenticular galaxies in our study were selected (\ref{sec:sample_selection}). We also consider a number of physical properties that most likely act as diagnostics of S0 formation mechanisms, such as stellar mass (Section \ref{sec:stellar_mass}), kinematics (Section \ref{sec:kinematics}), stellar populations (Section \ref{sec:ssp}),, and bulge-to-total luminosity ratio. The latter is retrieved from \citet{Simard+11}, except for the 21 galaxies in the pilot sample \citep{Coccato+20}, where we performed a parametric bulge-fit decomposition on the available R band images.
The properties of the environment of our sample galaxies are described in Section \ref{sec:environment}.

In total, we collected 994 lenticular galaxies. However, only a small sub-set of them have all the measured quantities that are needed in this study. Lenticular galaxies for which it was possible to measure kinematic parameters and "global" environment properties total 610. Lenticular galaxies for which it was possible to measure stellar kinematic parameters and "local" environment properties total 670. Lenticular galaxies for which it was possible to measure stellar kinematic, morphological parameters, stellar mass, bulge-to-disk ratio, and stellar populations total 329. Depending on the scope of a certain analysis, we used these different sub-samples.

The measured properties are listed on Table \ref{tab:sample} for a subset of targets. The full table is available in the electronic version of the paper online.

\subsection{Sample selection}
\label{sec:sample_selection}
Our galaxy sample includes lenticular galaxies from the MaNGA and SAMI\footnote{Sydney-AAO Multi-object Integral field spectrograph, \citet{Bryant+15, Croom+21}} galaxy surveys, plus the sample of our pilot study \citep{Coccato+20}.
Their morphological classifications are taken from the MaNGA Visual Morphology Catalogue (version 1.0.1\footnote{\url{https://data.sdss.org/datamodel/files/MANGA_MORPHOLOGY/manga_visual_morpho/}}, \citealt{Vazquez-Mata+22}) and from the SAMI visual morphology catalogues DR3\footnote{\url{https://datacentral.org.au/services/schema/\#sami.dr3.catalogues.other.VisualMorphologyDR3}} \citep{Cortese+16}. 

The MANGA visual morphological classification was carried out by judging in different steps mosaic images containing a  logarithmic-scaled r-band image, a filter-enhanced r-band image, the gri colour composite image from SDSS, a grz colour composite image, and a residual image after subtraction of a best surface brightness modelling from images in the Dark Energy Spectroscopic Survey \citep{Dey+19}. 
The SAMI visual morphological classification was performed separately by a number of astronomers either on SDSS or the VLT Survey Telescope RGB images by following two steps. First, an early/late type distinction was performed according to the presence or absence of disk, spiral arms, or signatures of star formation. Subsequently, the presence of distinct structural components (i.e., bulge/disk) in the galaxy was visually determined. Early-types with just a bulge were classical elliptical systems, whereas early-types with disks are S0s. Robustness of classification based on SDSS and VST images was also tested.
For objects not present in the  latter catalogues, we used the GAMA Visual morphology catalog (version 03\footnote{\url{http://www.gama-survey.org/dr3/data/cat/VisualMorphology/v03/}}). 

We used visual morphology classification to better compare to similar studies in the literature (e.g., \citealt{Rizzo+18, Coccato+20, Deeley+20}). Potential biases that can be introduced by visual classification will be discussed in Appendix \ref{app:selection_biases}, where we repeat the analysis by adding constraints on kinematics and specific star formation rate in our sample selection criteria.

\subsection{Stellar mass and star formation rates}
\label{sec:stellar_mass}

Stellar masses and star formation rates (SFRs) were taken from the $GALEX$-Sloan-$WISE$ Legacy Catalogue-2 \citep[GSWLC-2;][]{Salim+16, Salim+18} using a sky match with maximum allowed separation of 2$^{\prime\prime}$. In that catalogue, Stellar masses and SFRs were determined via SED fitting using the Code Investigating GALaxy Emission \citep[CIGALE;][]{Noll+09,  Boquien+19}. The photometry included in the fits was sourced from the Wide-Field Survey Explorer ($WISE$) in the mid-infrared, Sloan Digital Sky Survey in the optical, and the Galaxy Evolution Explorer ($GALEX$) in the UV. 

\begin{figure*}
\includegraphics[width=18cm]{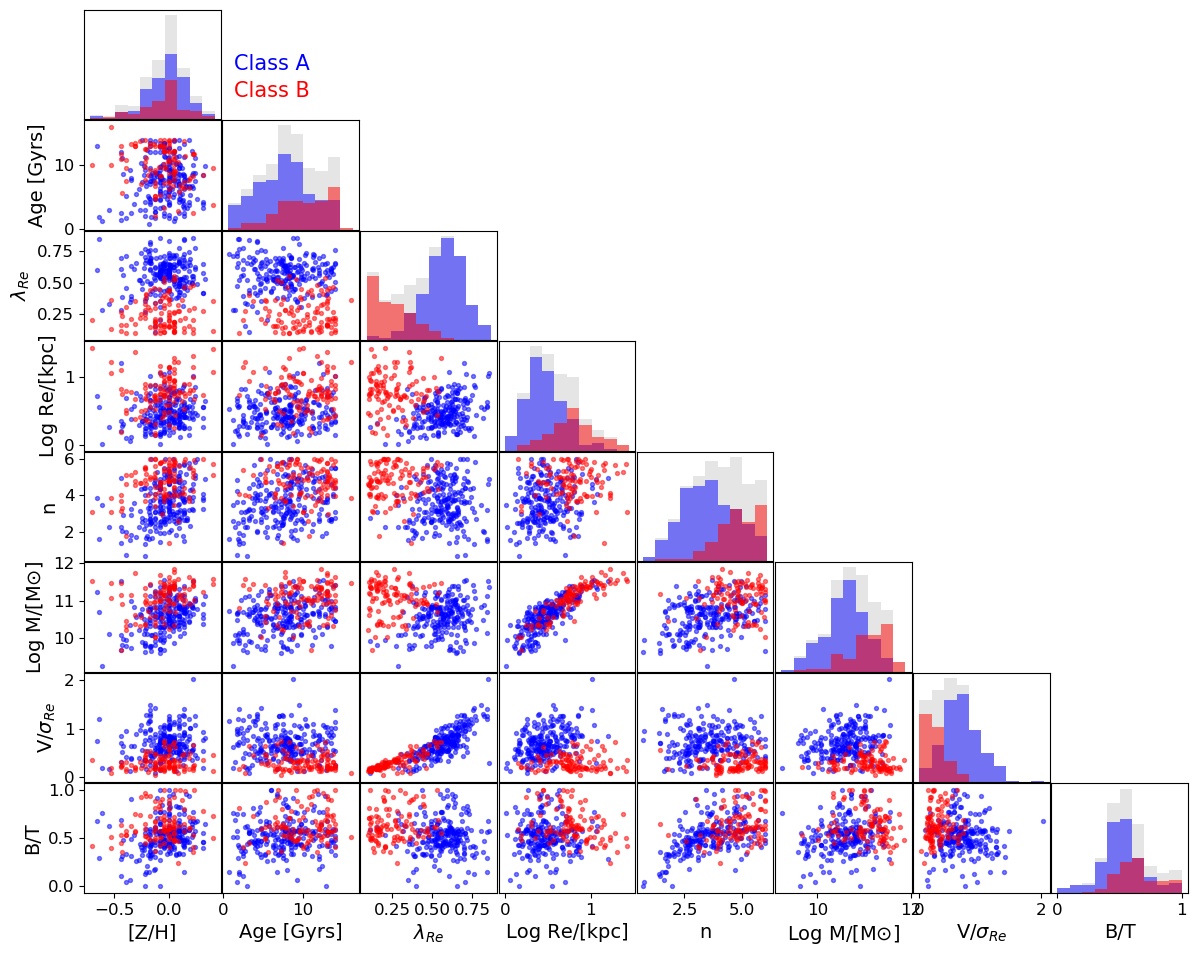}
\caption{Distribution of S0s in the parameter space defined by age, metallicity, angular momentum $\lambda$ and \vsigma\, effective radius $R_e$, S\`ersic index $n$, stellar mass $M$, and star formation rate SFR. Blue points indicate galaxies that belong to class A (associated to the "merger" scenario), red points indicate galaxies that belong to class B (associated to the "faded spiral" scenario). The number of S0s for which it was possible to determine the membership class is 329 (225 S0s in Class A, 104 in Class B). See text (Section\ref{sec:clustering}) for details on how the class association is done. The histograms indicate the distribution of a given parameter; grey histograms represent the total distribution, blue and red histograms indicate those of class A and B, respectively. 
}
\label{fig:clustering}
\end{figure*}

\subsection{Kinematics}
\label{sec:kinematics}
We recomputed the kinematic properties of the sample galaxies from the available 2-dimensional maps of flux, velocity, and velocity dispersion, to ensure homogeneity on how these quantities are computed. 
The specific angular momentum parameter within 1 effective radius, \lambdar\, was calculated following the method of \citet{Fraser+21}. Briefly, it is defined by \citet{Emsellem+07, Emsellem+11} as:

\begin{equation}
\lambda(R_e)= \frac{\langle R  |V|\rangle}{\langle R  \sqrt{V^{2} + \sigma^{2}}\rangle} = \frac{\sum_{i=0}^{N} F_{i}R_{i}|V_{i}|}{\sum_{i=0}^{N} F_{i}R_{i}\sqrt{V^{2}_{i} + \sigma^{2}_{i}}},
\end{equation}

where $F$ is the flux, $V$ the stellar rotational velocity, and $\sigma$ the stellar velocity dispersion of the $i-$th spaxel. In the same manner as \citet{Fraser+21}, we define $R$ as the semi-major axis of an ellipse on which spaxel $i$ lies. $N$ is the total number of spaxels within $1 R_e$.
Measurements are then corrected for inclination following \citet{Emsellem+11}.

The ratio of ordered to disordered motion within a galaxy, \vrotsigma, was calculated following \citet{Cortese+14}. Briefly, the amplitude of rotation is defined 
as the difference between the 90th and 10th percentile points of the stellar velocity histogram within $1 R_e$ ($W_{90,10}=V_{90}-V_{10}$). Then the velocity dispersion was computed on the galaxy integrated spectrum within $1 R_e$.

\begin{equation}
    V_{ROT}/\sigma  =  \frac{W_{90,10}}{2\left(1+z\right)\sin{\phi} \cdot \sigma(1R_e)},
\end{equation}

where $\phi$ is the inclination and $z$ the redshift. Inclinations are determined from the minor-to-major axis ratio following Equation 2 in \citet{Cortese+14} assuming an intrinsic axial ratio of 0.2 as in \citet{Catinella+12}.

\lambdar\ and \vrotsigma\ parameters are influenced by both the FWHM\footnote{Full width at half maximum.} of the PSF\footnote{Point spread function.} (the `seeing’ of the observation) and the inclination of the galaxy. We therefore correct them using the PSF correction of \citet{Harborne+20}.
Note that \lambdar\ is a flux-weighed quantity, whereas only the velocity dispersion is flux-weighted in the definition of \vrotsigma, not the rotational velocity.

\subsection{Stellar populations}
\label{sec:ssp}

Integrated light-weighted ages and metallicities were determined
by fitting the observed spectra (integrated over an aperture of 1 $R_e$ with the penalised pixel fitting method (\textsc{pPXF}, \citealt{Cappellari+04, Cappellari+17}) using simple stellar population  models from the MILES stellar library \citep{Vazdekis+10}. Emission lines and stellar continuum were fitted simultaneously by creating a set of Gaussian emission line templates to fit at the same time as the stellar templates. In this manner, only the intermediate region of the spectrum between the SAMI red and blue arm wavelength coverage regions (5760--6400 \AA) was masked. Barring the masked middle region, the spectrum was fit in the rest wavelength range 3900--6800~\AA. Regularisation was employed for the \textsc{pPXF} fit and the template normalisation for light-weighted properties performed over the range 5000--5500~\AA.

\subsection{Characterisation of the environment}
\label{sec:environment}

In this work we distinguish between "global environment", which we parametrise by the total halo mass of the group/cluster a galaxy belongs to, and "local environment", which we parametrise by the density of galaxies in a given area. 
The first step in characterising the environment of each galaxy is to determine whether or not it is isolated. To this purpose, we cross-match each galaxy in our sample with three catalogues: The "GAMA Galaxy Group Catalogue" \citep[version 10]{Robotham+11}, the "Groups and clusters of galaxies in the SDSS" \citep[version 12 June 2012]{Tempel+12}, generated using a friends-of-friends (FoF) based grouping algorithm, and the "2MASS Isolated Galaxies Catalogue" \citep[version 25 March 2011]{Karachentseva+10}. We classified galaxies in these catalogues that have no companions as "field galaxies". We classified galaxies in these catalogues that have at least 1 companion as "non-isolated galaxies". The 194 galaxies that are not in these catalogues, and therefore for which it is not possible to characterise their environment, are not considered in this analysis.
For all "non-isolated galaxies",  we used the total halo mass of the group/cluster  $M_{env}$ and the local density $\rho_5$ to characterise the environment.  

The global environment host halo mass and its error are computed following the prescription by \citet[section 4, equations 18 and 20]{Robotham+11}. Briefly, the method relies on the determination of the group velocity dispersion, its projected radius, and a scaling factor that depends on the number of galaxies in the group and the group redshift. Then, following  \citet{Deeley+20}, we classified the global environment on the basis of its total mass $M_T$. We defined as "small" group those environments that contain $N \leq 4$ galaxies and total mass $\log M_T/\msun \leq 13.5 $. We defined "big" group those environments that contain $N \leq 4$ galaxies and total mass $13.5   < \log M_T/\msun   \leq 14.5 $. Finally, we define cluster those environments with total mass $\log M_T/\msun> 14.5$. 

The local density is derived as the distance from the 5-th neighbour, $D_5$. This quantity is obtained from the "{\tt gama\_dr2.EnvironmentMeasure} catalog" (GAMA data release 2, catalogue version 5, \citealt{Brough+13}) and the "Galaxy Environment for MaNGA Value Added Catalog" version 1.0.1 (\citealt{Aguado+19}, Argudo-Fernandez et al, in prep.).

\begin{landscape}
\begin{table}
\caption{Measured properties of the sample galaxies.}
\begin{scriptsize}
\begin{tabular}{lccccccccccccccccc}
\hline
ID          & Dataset    &Dist.  &Log $M^\star$       &	     Age             &	     [Z/H]               &      \lambdar     &    \vrotsigma    &      Log M\_env    & Ngal  & 	 Dens5     &Re     &$n$    &     $B/T$       &       Log(SFR)               &Class   & Dist$_A$  & Dist$_B$ \\
  (1)       &  (2)       & (3)   &  (4)               &          (5)         &        (6)                &       (7)         &        (8)       &        (9)         &  (10) &    (11)       &  (12) & (13)  &     (14)        &         (15)                 &(16)    &  (17)     &   (18)   \\
            &            &Mpc    &Log$(/[M_{\odot}])$ &       Gyr            &                           &                   &                   & Log$(/[M_{\odot}])$&       &   Mpc$^{-2}$    & kpc   &       &	             &  Log$(/[M_{\odot} yr^{-1}$]) &        &           &          \\
\hline
7992	    & SAMI	 & 346.8 & $11.09 \pm 0.03$   & $11.4^{+0.2}_{-0.5}$ &	 $0.14^{+0.03}_{-0.01}$  & $ 0.22 \pm 0.03 $ &	$0.23 \pm 0.03$ &  $0.00  \pm 0 $    &	2  & NA            & 5.30  & 5.96  & $0.99 \pm 0.01$ &	 $-0.447 \pm  0.716 $       &   B    &   0.919   &  0.473   \\
9062	    & SAMI	 & 93.4  & $10.25 \pm 0.02$   & $7.3^{+0.2}_{-0.5}$  & 	 $0.00^{+0.01}_{-0.01}$  & $ 0.33 \pm 0.04 $ &	$0.31 \pm 0.04$ &  $0.00  \pm 0 $    &	2  & NA            & 1.98  & 5.77  & $0.63 \pm 0.01$ &	 $-1.685 \pm  0.651 $       &   B    &   0.600   &  0.543   \\
16317	    & SAMI	 & 146.5 & $10.59 \pm 0.01$   & $4.7^{+1.8}_{-1.5}$  & 	$-0.01^{+0.05}_{-0.06}$  & $ 0.44 \pm 0.05 $ &	$0.55 \pm 0.07$ &  $10.59 \pm 0.01$  &	1  & NA            & 3.97  & 5.16  & $0.34 \pm 0.01$ &	 $-2.750 \pm  0.480 $       &   A    &   0.450   &  0.583   \\
16926	    & SAMI	 & 215.4 & $10.34 \pm 0.07$   & $11.5^{+2.5}_{-4.9}$ & $-0.44^{+0.12}_{-0.04}$   &  NA              &	NA              &  $10.34 \pm 0.07$  &	1  & NA            & 2.58  & 1.52  & $0.23 \pm 0.01$ &	 $-0.226 \pm  0.095 $       &   U    &     U     &     U    \\
31448	    & SAMI	 & 347.4 & $10.91 \pm 0.03$   & $8.1^{+0.0}_{-4.8}$  &	 $0.19^{+0.09}_{-0.03}$  & $ 0.46 \pm 0.06 $ &	$0.51 \pm 0.06$ &  $13.43 \pm 0.7$   &	4  & NA            & 6.53  & 5.53  & $0.61 \pm 0.01$ &	 $-2.009 \pm  0.584 $       &   B    &   0.540   &  0.427   \\
47293	    & SAMI	 & 224.8 & $10.29 \pm 0.04$   & $7.9^{+0.0}_{-4.8}$  & 	$-0.08^{+0.03}_{-0.09}$  & $ 0.63 \pm 0.08 $ &	$0.73 \pm 0.09$ &  $13.39 \pm 0.6$   &	8  & $2.0 \pm 0.8 $& 2.41  & 3.07  & $0.27 \pm 0.01$ &	 $-1.292 \pm  0.951 $       &   A    &   0.308   &  0.845   \\
78634	    & SAMI	 & 237.1 & $10.41 \pm 0.03$   & $9.8^{+1.0}_{-2.6}$  & 	$-0.03^{+0.05}_{-0.06}$  & $ 0.55 \pm 0.07 $ &	$0.50 \pm 0.06$ &  $0.00  \pm 0$     &	3  & $1.09\pm 0.10$& 2.86  & 2.70  & $0.43 \pm 0.02$ &	 $-1.131 \pm  0.704 $       &   A    &   0.264   &  0.686   \\
91686	    & SAMI	 & 233.3 & $10.97 \pm 0.03$   & $9.4^{+0.3}_{-1.6}$  &   $0.02^{+0.03}_{-0.05}$  & $ 0.64 \pm 0.07 $ &	$0.83 \pm 0.10$ &  $13.27 \pm 0.6 $  &	9  & $2.7 \pm 0.3 $& 6.10  & 4.92  & $0.50 \pm 0.01$ &	 $-0.867 \pm  0.628 $       &   A    &   0.406   &  0.599   \\
98097	    & SAMI	 & 120.5 & $10.30 \pm 0.02$   & $7.5^{+0.4}_{-1.1}$  & 	 $0.09^{+0.01}_{-0.01}$  & $ 0.39 \pm 0.05 $ &	$0.40 \pm 0.05$ &  $0.00  \pm 0$     &	2  & NA            & 2.04  & 5.05  & $0.66 \pm 0.01$ &	 $-0.413 \pm  0.193 $       &   A    &   0.470   &  0.523   \\
1-246484    & MANGA	 & 130.0 & $ 9.91 \pm 0.04$   & $3.3^{+0.0}_{-0.3}$  & 	$-0.13^{+0.02}_{-0.01}$  & $ 0.11 \pm 0.01 $ &	$0.09 \pm 0.01$ &  $9.91  \pm 0.04$  &	1  & NA            & 1.82  & 2.78  & $0.59 \pm 0.08$ &	 $-2.236 \pm  0.250 $       &   A    &   0.819   &  0.847   \\
1-351694    & MANGA	 & 237.7 &       NA           & $13.0^{+0.0}_{-0.1}$ & $-0.17^{+0.01}_{-0.01}$   & $ 0.11 \pm 0.01 $ &	$0.14 \pm 0.02$ &  $0.00  \pm 0$     &	3  & $1.02\pm 0.08$& 4.32  & 6.00  & $0.72 \pm 0.01$ &	 NA                         &   U    &     U     &     U    \\
1-266074    & MANGA	 & 577.5 & $11.58 \pm 0.04$   & $9.5^{+0.0}_{-0.1}$  & 	 $0.40^{+0.01}_{-0.01}$  & $ 0.11 \pm 0.01 $ &	$0.13 \pm 0.02$ &  $0.00  \pm 0$     &	2  & $0.42\pm 0.01$& 25.61 & 5.70  & $0.50 \pm 0.03$ &	 $-1.133 \pm  0.792 $       &   B    &   1.164   &  0.712   \\
1-318364    & MANGA	 & 521.3 & $11.43 \pm 0.04$   & $9.5^{+0.1}_{-0.1}$  & 	 $-0.04^{+0.01}_{-0.01}$ & $ 0.18 \pm 0.02 $ &	$0.20 \pm 0.02$ &  $0.00  \pm 0$     &	3  & $0.47\pm 0.02$& 8.32  & 4.07  & $0.69 \pm 0.02$ &	 $-0.839 \pm  0.700 $       &   B    &   0.770   &  0.256   \\
1-113520    & MANGA	 & 72.2  & $ 9.95 \pm 0.03$   & $5.5^{+0.0}_{-0.3}$  & 	$-0.09^{+0.00}_{-0.01}$  & $ 0.20 \pm 0.02 $ &	$0.30 \pm 0.04$ &   NA               &	0  & $3.05\pm 0.70$& 1.05  & 2.27  & $0.53 \pm 0.01$ &	 $-1.594 \pm  0.906 $       &   A    &   0.721   &  0.890   \\
1-284335    & MANGA	 & 149.6 &  $9.79 \pm 0.05$   & $4.3^{+0.0}_{-0.3}$  & 	$-0.22^{+0.02}_{-0.01}$  & $ 0.27 \pm 0.03 $ &	$0.34 \pm 0.04$ &  $0.00  \pm 0$     &	2  & $1.22\pm 0.11$& 2.44  & 4.02  & $0.42 \pm 0.03$ &	 $-1.423 \pm  0.313 $       &   A    &   0.614   &  0.736   \\
1-633000    & MANGA	 & 84.8  & $10.08 \pm 0.04$   & $1.5^{+0.0}_{-0.1}$  &  $-0.44^{+0.00}_{-0.01}$  & $ 0.29 \pm 0.03 $ &	$0.30 \pm 0.04$ &  $10.08 \pm 0$.04  &	1  & NA            & 1.58  & 2.28  & $0.69 \pm 0.09$ &	 $ 0.012 \pm  0.201 $       &   A    &   0.800   &  0.978   \\
1-178695    & MANGA	 & 302.2 & $11.16 \pm 0.03$   & $13.0^{+0.1}_{-0.1}$ & $-0.04^{+0.00}_{-0.01}$   & $ 0.30 \pm 0.04 $ &	$0.38 \pm 0.05$ &   NA               &	0  & NA            & 5.09  & 5.31  & $0.73 \pm 0.01$ &	 $-0.746 \pm  0.873 $       &   B    &   0.712   &  0.260   \\
2MIG 131    & PaperI     & 106.5 & $11.46 \pm 0.18$   & $16.0^{+0.5}_{-0.5}$ & $-0.53^{+0.02}_{-0.02}$   & $ 0.36 \pm 0.04 $ &	$0.09 \pm 0.01$ &  $11.46 \pm 0.28$  &	1  & NA            & 16.63 & 3.85  & $0.51 \pm 0.10$ &	NA                          &   B    &   1.033   &  0.729   \\
2MIG 1546   & PaperI     & 64.0  & $11.13 \pm 0.01$   & $8.4^{+0.5}_{-0.5}$  &  $-0.17^{+0.03}_{-0.03}$  & $ 0.60 \pm 0.07 $ &	$0.33 \pm 0.04$ &  $11.13 \pm 0.013$ &	1  & NA            & 7.26  & 2.74  & $0.44 \pm 0.10$ &	NA                          &   A    &   0.450   &  0.648   \\
CCC 43	    & PaperI     & 48.7  & $10.96 \pm 0.02$   & $13.2^{+0.6}_{-0.6}$ & $-0.13^{+0.04}_{-0.04}$   & $ 0.46 \pm 0.06 $ &	$0.19 \pm 0.02$ &  $14.54 \pm 0.50$  &	50 & NA            & 5.21  & 4.98  & $0.56 \pm 0.10$ &	NA                          &   B    &   0.598   &  0.370   \\
CCC 122	    & PaperI     & 48.7  & $10.88 \pm 0.01$   & $11.3^{+0.2}_{-0.2}$ & $-0.04^{+0.01}_{-0.01}$   & $ 0.62 \pm 0.07 $ &	$0.44 \pm 0.05$ &  $14.54 \pm 0.50$  &	50 & NA            & 4.34  & 2.47  & $0.48 \pm 0.10$ &	NA                          &   A    &   0.388   &  0.689   \\
\hline
\end{tabular}
\end{scriptsize}
\label{tab:sample}
\begin{minipage}{23.5cm}
Note: Measured quantities for the sample galaxies. The full table is available on-line as supplementary material in the on-line version of the paper. The quantities and errors (whenever available) are derived as detailed in Section \ref{sec:sample}.\\
Col.  1: Galaxy ID.\\
Col.  2: Source dataset.\\
Col.  3: Distance in Mpc.\\
Col.  4: Stellar mass and its error, according to \citet{Salim+16, Salim+18} (see Section 2.2), in Log$_{10}$ of solar masses.\\
Col.  5: Age of the stellar population and its upper and lower errors in Gry, computed as described in Section 2.4.\\
Col.  6: Metallicity of the stellar population and its upper and lower errors, in Log$_{10}$ of solar units, computed as described in Section 2.4.\\
Col.  7: Lambda parameter within 1 effective radiuss and its error, computed as described in Section 2.3.\\
Col.  8: Ratio between the stellar rotational velocity and stellar velocity dispersion, computed as described in Section 2.3.\\
Col.  9: Mass of the host environment the galaxy belongs to and its error, in Log$_{10}$ of solar masses. For isolated system the value is a copy of Column 4. See Section 2.5 for further information on the reference source.\\
Col. 10: Number of galaxies in the system the galaxy belongs to. See Section 2.5 for further information on the reference source.\\
Col. 11: Number density of galaxies and its error computed within the 5th neighbour in units of Mpc$^{-2}$. See Section 2.5 for further information on the reference source.\\
Col. 12: Effective radius in units of kpc, according to  Galaxy And Mass Assembly (GAMA) input catalogue (3rd release, catalogue name: InputCatGAMADR3.fits) or NASA Sloan Atlas DR13 (https://data.sdss.org/sas/dr13/sdss/atlas/v1/), if first not available.\\
Col. 13: n parameter of the Sersic fit, according to NASA Sloan Atlas DR13 (https://data.sdss.org/sas/dr13/sdss/atlas/v1/).\\
Col. 14: Bulge-to-Total light ratio and its error, according to \citet{Simard+11} and \citet{Coccato+20}.\\
Col. 15: Star formation rate and its error, according to \citet{Salim+16, Salim+18} (see Section 2.2) in Log$_{10}$ of solar masses per year.\\
Col. 16: The class the galaxy belong to (A or B), according to the classication performed by the KMeans algorithm (see Section 3.1). Galaxies with un-assigned class are marked with U.\\
Col. 17: Distance in the multi-dimensional parameter space to the center of the class A, in normalized units (see Section 3.1). Galaxies with un-assigned class are marked with U.\\
Col. 18: Distance in the multi-dimensional parameter space to the center of the class B, in normalized units (see Section 3.1). Galaxies with un-assigned class are marked with U.
\end{minipage}
\end{table}
\end{landscape}

\section{Analysis}
\label{sec:analysis}

\begin{figure*}
\vbox{
\hbox{
 \includegraphics[width=6cm]{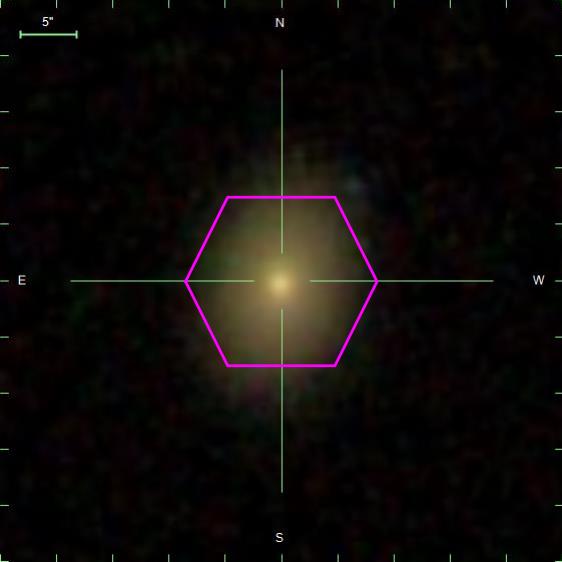}
 \includegraphics[width=6cm]{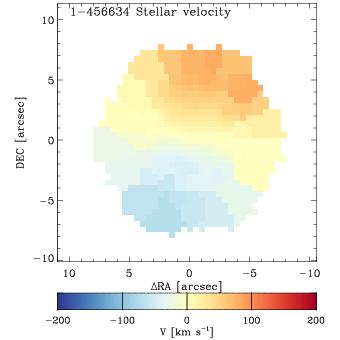}
 \includegraphics[width=6cm]{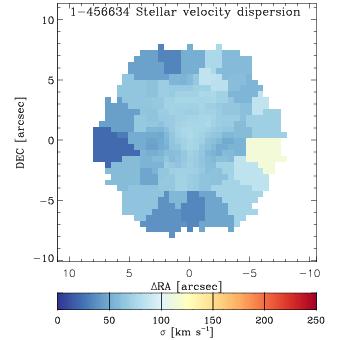}}
 \vspace{1cm}
 \hbox{
 \includegraphics[width=6cm]{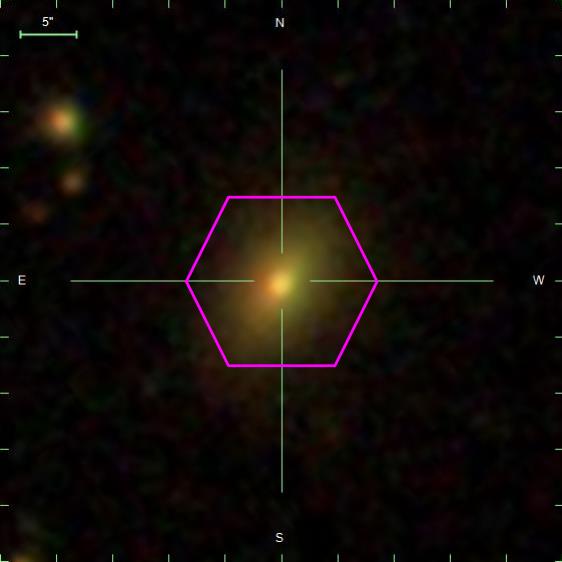}
 \includegraphics[width=6cm]{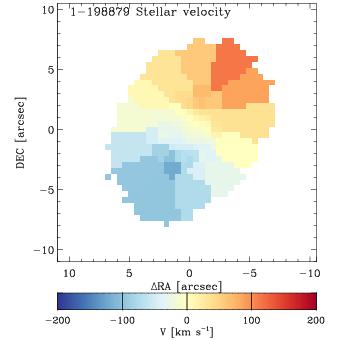}
 \includegraphics[width=6cm]{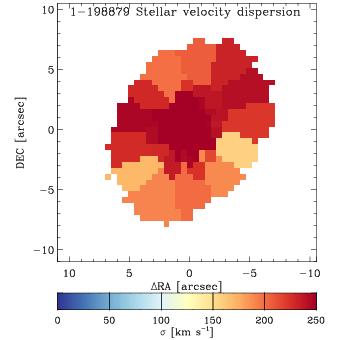}}
 }
\caption{Examples of two galaxies in our sample: 1-456634 (Class A) in the top panels, and 1-198879 (Class B) in the bottom panels. The SDSS images with the MANGA footprint (Image source: https://dr17.sdss.org/marvin/), the stellar velocity and velocity dispersion two-dimensional fields are shown in the left, central, and right-hand panels, respectively.}
\label{fig:sample_image}
\end{figure*}

\subsection{Two classes of lenticular galaxies}
\label{sec:clustering}
Driven by the idea that lenticular galaxies formed by different processes have different properties, we investigated and compared the distribution of the global properties. The goal is to understand if we can identify classes of objects with similar properties that we can associate to one of the two formation paths (faded spiral, or merging).

In Figure \ref{fig:clustering} we have compared the galaxy properties defined in Section \ref{sec:sample} to each-other. The figure shows also the histogram with the distribution of that specific property. The physical properties we considered are luminosity-weighted stellar age (Age) and metallicity ([Z/H]) within 1 effective radius, the effective radius (Log $R_e$/[kpc]), the S\`ersic parameter of the photometric decomposition ($n$, form a single component fit), the stellar mass (Log [$M^\star /M_\odot$]), the specific spin parameter (\lambdar), ratio of ordered and non-ordered motions (\vrotsigma) within $1 R_e$, and the bulge-to-total luminosity ratio (B/T).
Only the sample of 329 lenticular galaxies for which all the physical quantities are available were used in the analysis. 

The histograms of the various properties show bi-modal distributions, mainly in the Age, kinematics, and S\`ersic index $n$. Assuming the existence of two distinct populations of S0 galaxies, and in order to better associate each galaxy to one of the two,
we applied a simple KMeans clustering algorithm to the data. The KMeans algorithm clusters data by separating the sample into two groups of equal variance, minimising a criterion known as the "inertia" or "within-cluster sum-of-squares":
\begin{equation}
   \sum_{i = 0}^{n} min\left(|| x_i-\mu_j||^2\right)
\end{equation}
where $x_i$ is the value of the quantity $x$ of the $i$-th galaxy, and $\mu_j$ is the mean value of that quantity for the $j=1,2$ class or cluster. Each physical quantity was normalised so that its values range in the interval 0,1; this is to remove the physical units from the process and avoid that quantities with large value (in a given unit) weight more than others in the computation of the inertia. The computation was carried on with the Python implementation of KMeans in the scikit-learn module. The goal is to find an indication of the presence of two S0 sub-populations with a simple approach with the minimum number of free parameters.
Other, more sophisticated clustering techniques (e.g., Principal Components Analysis, populations with different variance) and the assumption of more than 2 populations might be explored in further follow-up.

The KMeans algorithm groups together galaxies that have similar properties in the in the multi-parameter space defined in Fig. \ref{fig:clustering} and flags them accordingly into two groups, or classes. A galaxy belongs to one group or another depending on the distance to the group center defined in the N-dimensional parameter space. The coordinates of the centers of the two classes are shown in Table \ref{tab:centers}; their errors are estimated via bootstrapping. The ratio of the distances to the two centres, $D_R$\footnote{$D_R$ = Dist$_A$ / Dist$_B$ for galaxies in Class B, and $D_R$ = Dist$_B$ / Dist$_A$ for galaxies in Class B.} , can be used as indication on how robust the class determination for a galaxy is. A galaxy with distance ratio $D_R=1$ has uncertain classification (i.e. it could belong to one class or another). The higher the ratio, the more robust the class determination is. In Table \ref{tab:sample} we report the distances to the centre of each group (in normalized coordinates). In our study, we checked that the results shown in Section \ref{sec:environment} do not change if we use a sub-sample with ratios larger than a threshold of 1.25. For higher thresholds, the trends shown in our analysis still hold but the error bars becomes bigger due to the small number of galaxies in the sub-sample.

The two groups defined by the clustering algorithm are distinct in the parameter space shown in Fig. \ref{fig:clustering}, in particular in the kinematic parameters. Fig. \ref{fig:sample_image} shows one typical example of S0 in each class.
%
%
For convenience, we name these two groups of S0s Class A and Class B. 
\begin{itemize}
\item {\bf Class A}. Lenticulars of this class are on average younger, more rotationally supported, smaller and with lower S\`ersic indices, less massive, and tend to have slightly smaller B/T (i.e., less luminous bulges at a given total luminosity).
\item {\bf Class B}. Lenticulars of this class are on average older, more pressure supported, larger and with larger S\`ersic index, more massive, and tend to have slightly higher B/T (i.e., more luminous bulges at given total luminosity).
The two classes have similar stellar metallicity.
\end{itemize}

The clustering is an outcome of the existence of correlations between different parameters (e.g., mass and size or, B/T and stellar velocity dispersion). A null result, i.e. the existence of a single population, would have shown a separation only in one parameter, whereas blue and red points would have occupied the same regions in the panels of the other parameters. 
Although there is not a net distinction nor threshold in the properties of galaxies formed via the two scenarios, nor a one-to-one correspondence between a given class and a formation scenario, we can recognise (see Section \ref{sec:intro}) that S0s in Class A have properties that are more in common with being formed through the "faded-spiral" scenario. On the other hand, S0s in Class B share similar properties of merger products. We therefore {\it assume} this association between S0 class and formation scenario through the rest of the paper.
\begin{centering}
\begin{table}
\caption{Class centers.}
\begin{tabular}{l c c}
\hline
Parameter     & Class A & Class B \\
\hline
[Z/H]         & $-0.03 \pm 0.03 $ & $-0.06  \pm 0.02 $\\               
Age [Gry]     & $ 7.6  \pm 1.3  $ & $10.0   \pm 1.2  $\\               
\lambdar      & $ 0.57 \pm 0.15 $ & $ 0.26  \pm 0.15 $\\               
LogM/M$\odot$ & $10.58 \pm 0.27 $ & $11.06  \pm 0.27 $\\               
$R_e$ [kpc]   & $1.49  \pm 0.16 $ & $ 1.79  \pm 0.16 $\\               
n             & $3.56  \pm 0.64 $ & $ 4.76  \pm 0.65 $\\               
\vsigma$(Re)$ & $0.71  \pm 0.19 $ & $ 0.30  \pm 0.20 $\\               
B/T           & $0.52  \pm 0.07 $ & $ 0.64  \pm 0.07 $\\                
\hline
\label{tab:centers}
\end{tabular}
\end{table}
\end{centering}

\subsection{Dependency of formation mechanisms with environment}
\label{sec:formation_and_environment}

In Section \ref{sec:clustering} we collected evidence of two main classes (or sub-populations) of lenticular galaxies. Encouraged by the similarities between physical properties of the galaxies in each class and the properties of the two main formation pathways, we assumed a connection between formation path and galaxy sub-population. According to this assumption, galaxies in class A are formed via a faded-spiral scenario, galaxies in class B are formed via merger events.
In this Section we investigate the distribution of different galaxy sub-populations in different environments, with the goal to highlight a connection between environment and formation mechanism.

\begin{figure}
\psfig{file=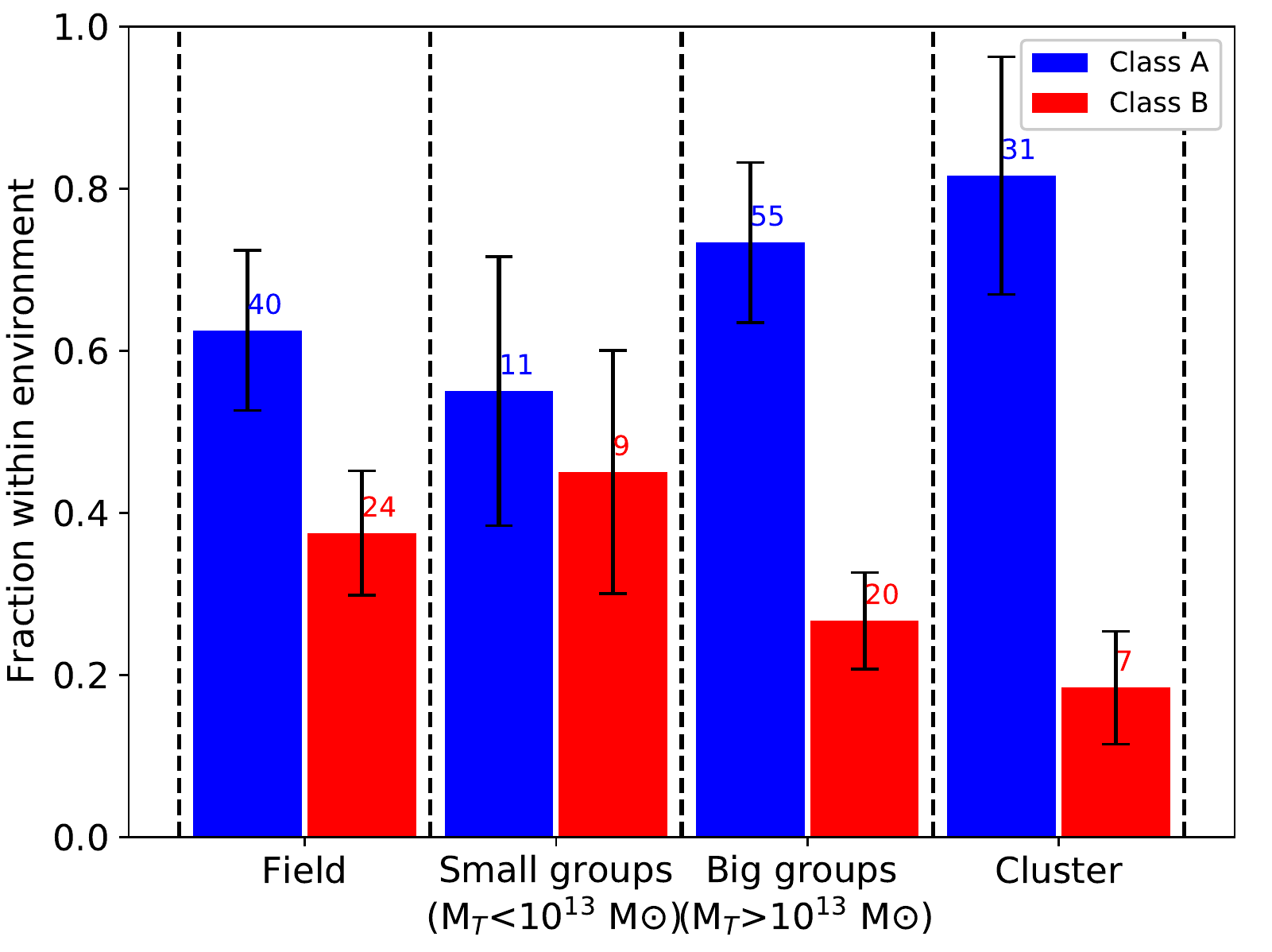,clip=,width=9cm}
\caption{Fraction of galaxies of different classes in each environment. The fractions are normalised so that their sum in each environment is 1.  Error bars are computed assuming Poisson statistics on the number of S0s in each environment. Dashed lines identify the density bins. The number of galaxies contributing to each bin is indicated at the top of each bar. The number of S0s used in this analysis is 197. 
The same trend is observed if we considered only the sub-sample of galaxies for which the ratio of distances to the class centre in the multidimensional parameter space is higher than 1.25 (see Section \ref{sec:clustering} for details). In this case the sample of galaxies goes down to 152.}
\label{fig:env_mass}
\end{figure}

Figure \ref{fig:env_mass} shows the fraction of S0 classes in different environments: field, small groups , big groups, and clusters. Class A, which includes S0 galaxies associated to the "faded spiral" formation scenario, is the most common class. This indicates that the faded spiral scenario (or to be more precise, all the formation mechanisms that include consumption or removal of gas, and AGN feedback) is the most efficient mechanism in forming lenticular galaxies, at least for the objects in our sample. Moreover, the efficiency of the "faded spirals" scenario is higher in high mass groups and clusters. The merger scenario, associated to S0s in class B, is also a viable mechanism, and becomes more efficient in smaller groups. 

\begin{figure}
\hspace{-.5cm}
\psfig{file=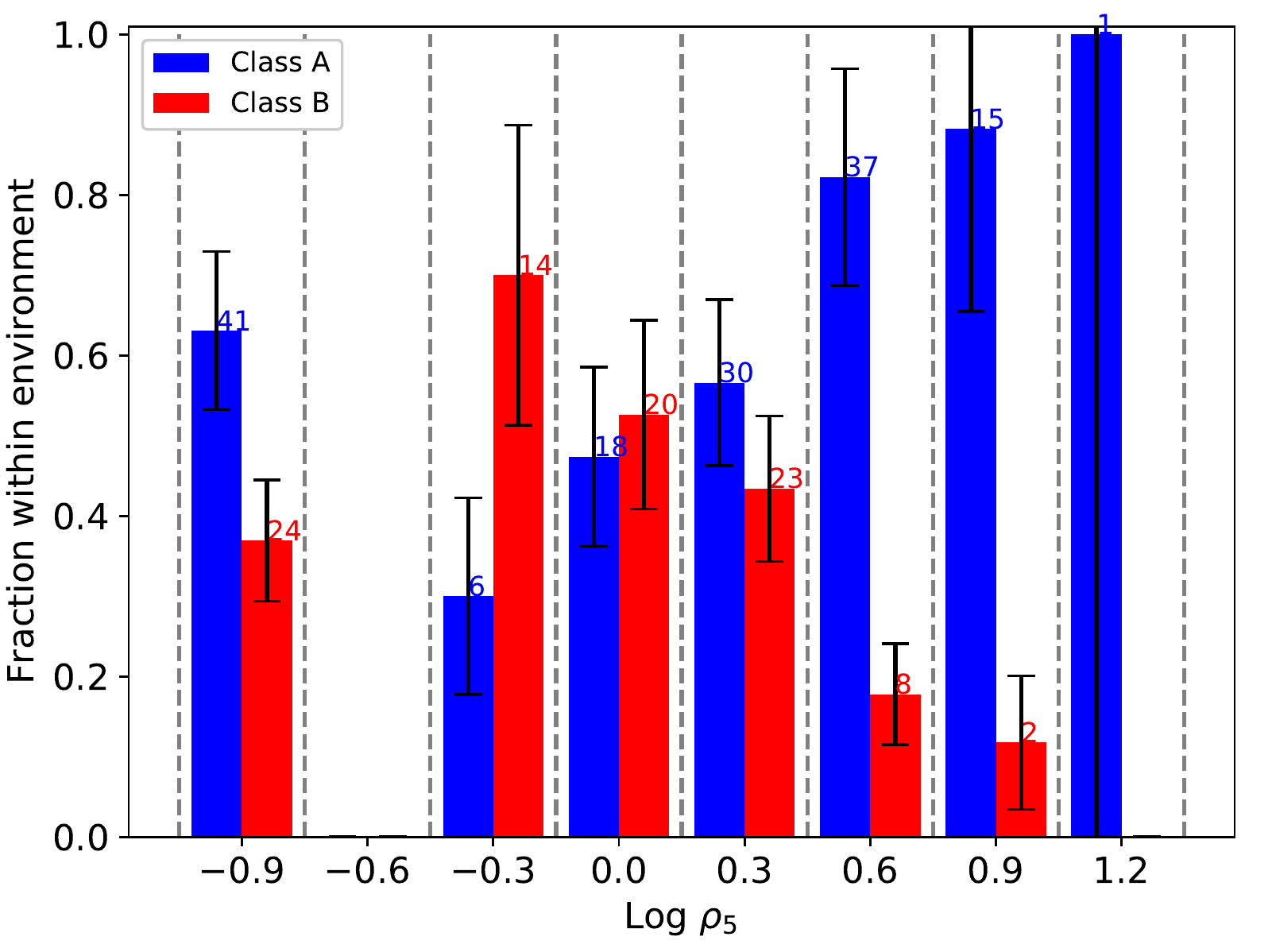,clip=,width=9cm}
\caption{Fraction of galaxies of different classes for each bin of density $\rho_5$. Field galaxies for which the density is undefined, are artificially shown in the left-most bin. Error bars are computed assuming Poisson statistics on the number of S0s in each environment. The number of galaxies contributing to each bin is indicated at the top of each bar. The number of galaxies used in this analysis is 239. The same trend is observed if we considered only the sub-sample of galaxies for which the ratio of distances to the class centre in the multidimensional parameter space is higher than 1.25 (see Section \ref{sec:clustering} for details). In this case the sample of galaxies goes down to 189.}
\label{fig:density1}
\end{figure}

Figure \ref{fig:density1} shows the fraction of S0 sub-populations as a function of local density, measured from the distance to the 5-th nearest neighbour (Section \ref{sec:sample}). 
Consistently with the results shown in Fig. \ref{fig:env_mass}, the use of local density as metric for the environment reveals that the faded-spiral is the most efficient scenario and its efficiency increases with environmental density. For very low density environments (but not for field galaxies) there is an indication that the merger scenario is the most efficient mechanism.

Our work relies on the combination of data coming from two different surveys, MaNGA and SAMI. These surveys are different in terms of depth, instrument set-up, and target environments. We therefore checked the validity of our results if considering data from only one survey at time. Both surveys clearly identify two classes of lenticular galaxies, with properties consistent with the two families of scenarios. With respect to the frequency of each class in a given environment, the results are confirmed when using the MaNGA data only, but are more noisy when using SAMI galaxies alone. This is because the sample of SAMI galaxies for which it is possible to have a clear separation between class A and B (115 galaxies) is  smaller than the MaNGA (195 galaxies), therefore the relation with environment is less pronounced.
The relation between physical properties and environment, however, holds also for individual surveys if we consider kinematic properties only, as seen in previous studies (see Section \ref{sec:comparison}).

\begin{figure}
\hspace{-.5cm}
\psfig{file=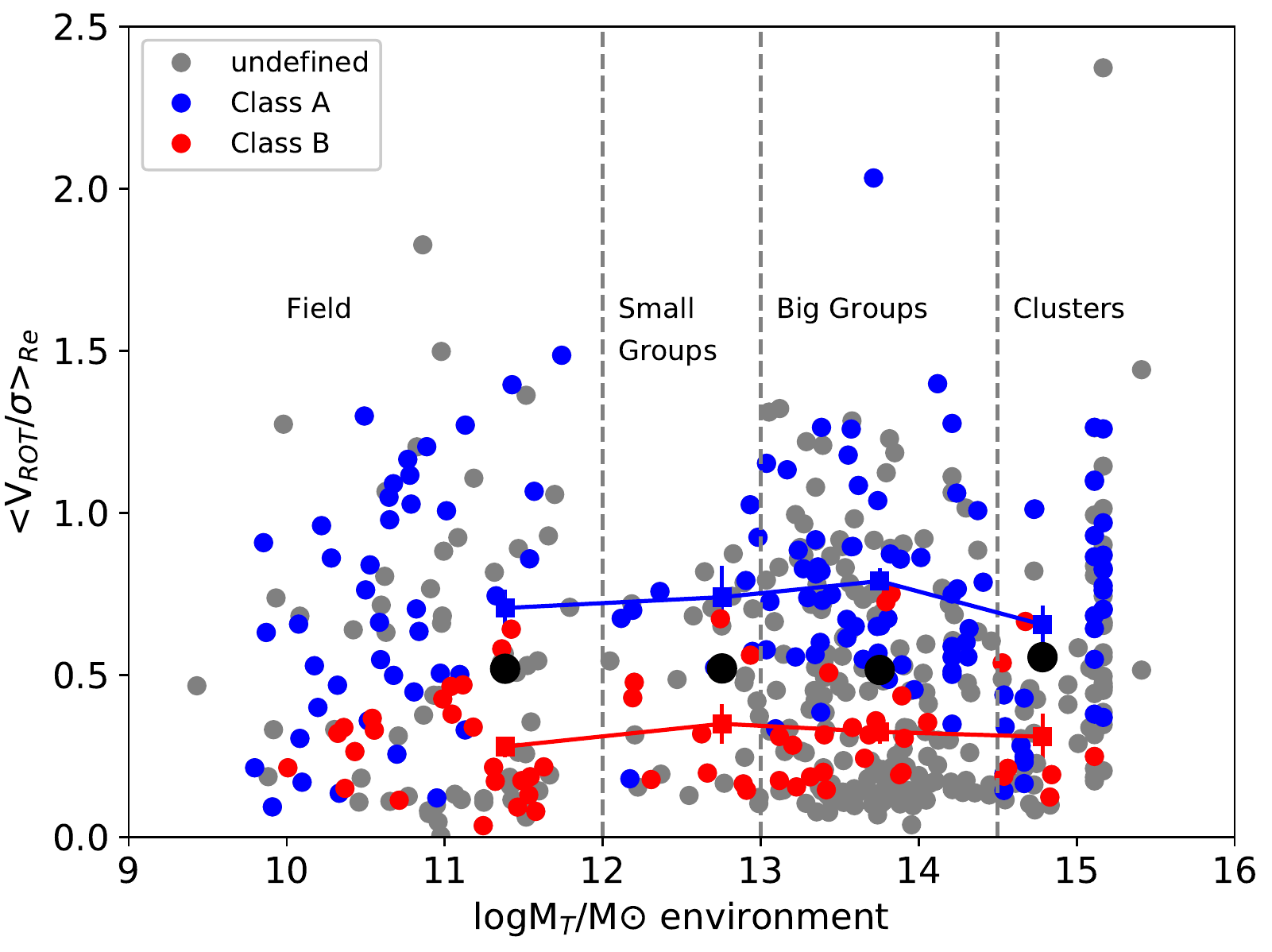,clip=,width=9cm}
\caption{Distribution of \vrotsigma\ as a function of total mass of the environment. For galaxies in the field, the mass is the total mass of the galaxy. Black circles are the mean values for each environment. The blue and red lines indicate the trend of the averages of the individual S0 classes. The number of objects for which it was possible to measure \vrotsigma\ and global environmental mass is 610. Those for which it was also possible to determine the class are 197. The same trend is observed if we considered only the sub-sample of galaxies for which the ratio of distances to the class centre in the multidimensional parameter space is higher than 1.25 (see Section \ref{sec:clustering} for details). In this case the sample of galaxies goes down to 152.}
\label{fig:env_mass3}
\end{figure}

\begin{figure}
\hspace{-.5cm}
\psfig{file=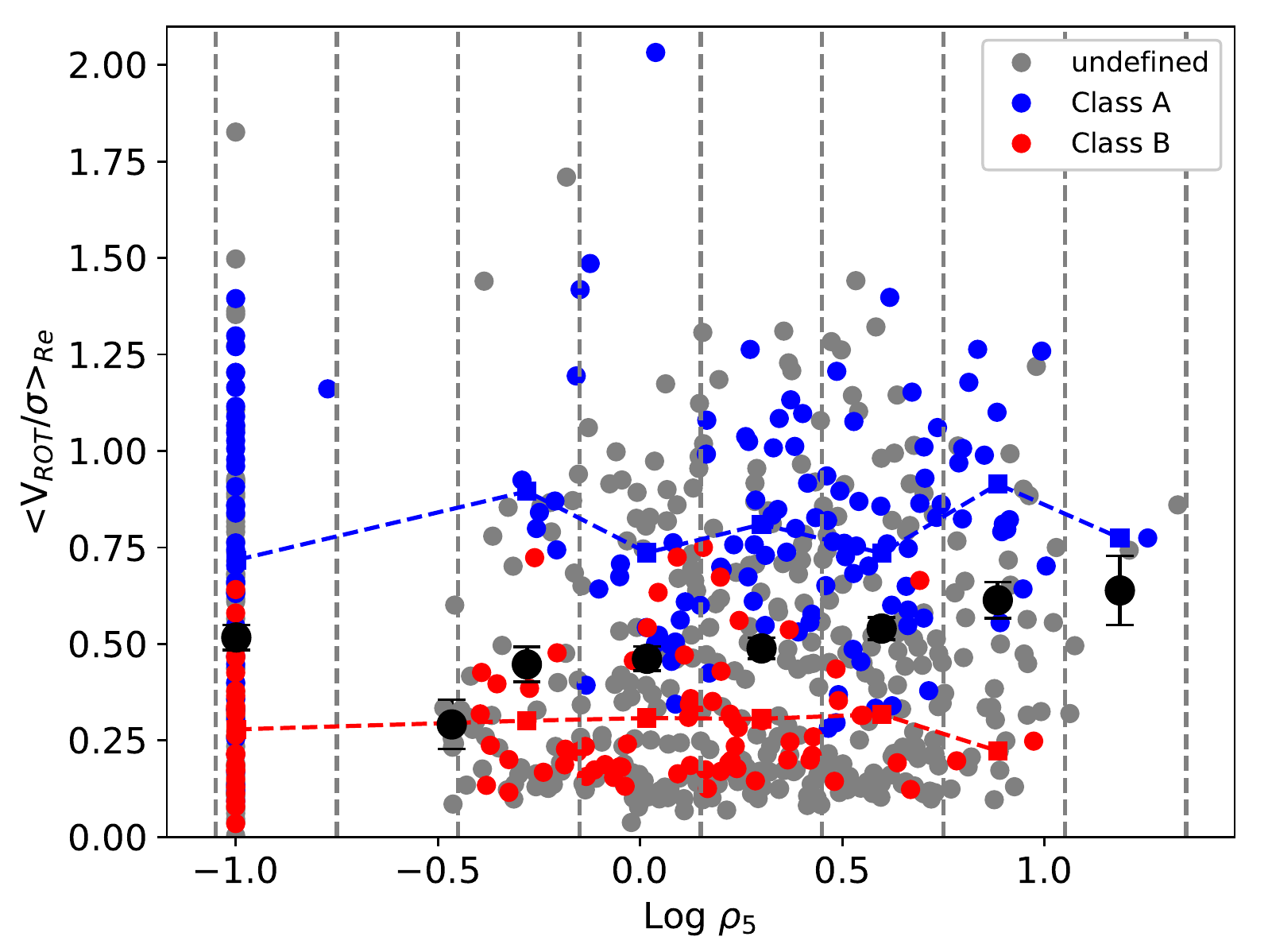,clip=,width=9cm}
\caption{Distribution of \vrotsigma\ as a function of local density. Galaxies in the field are artificially set to $\rho_5 = -1$. Black circles are the mean value for each density bin. The blue and red lines indicate the trend of the averages in each bin for the two S0 classes. The number of S0s for which it was possible to measure \vrotsigma, for which $\rho_5$ was available is 670. Those for which it was also possible to determine the class are 239. The same trend is observed if we considered only the sub-sample of galaxies for which the ratio of distances to the class centre in the multidimensional parameter space is higher than 1.25 (see Section \ref{sec:clustering} for details). In this case the sample of galaxies goes down to 189.}
\label{fig:density2}
\end{figure}

In order to connect our work with previous studies that used kinematics alone as a proxy for S0 formation mechanism (pressure supported systems formed via mergers, more rotationally supported systems via transformation of spiral progenitors, e.g., \citealt{Coccato+20, Deeley+20}), we show in Figures \ref{fig:env_mass3} and \ref{fig:density2} the dependency of \vrotsigma\ with environment.

Fig. \ref{fig:env_mass3} shows the value of \vrotsigma\ as a function of host-halo mass. There is not a significant trend with environment mass, despite galaxies in class A being more frequent in clusters and therefore that environment should be characterised by the kinematics of that S0 sub-population.
Fig. \ref{fig:density2} shows the value of \vrotsigma\ as a function of local density. Field lenticulars are shown at $\log \rho_5 =-1$ for convenience. The average \vrotsigma\ is computed in density bins and increases with increasing density. This is a consequence that the fraction of class A galaxies is higher in higher density bins (Fig. \ref{fig:density1}) and the \vrotsigma\ is higher for this sub-population (Fig. \ref{fig:clustering}).

\subsection{Comparison with literature}
\label{sec:comparison}

Our results mostly agree with previous studies where the formation pathways (merger or stripped spiral) were inferred only by looking at kinematic properties: \vrotsigma\ and $\lambda$ (both within $R_e$ and 1.5 $R_e$, as in \citealt{Coccato+20}) or $\lambda$ within 1.5 $R_e$ (as in \citealt{Deeley+20}). Both studies showed that the fraction of galaxies that are more rotationally supported (and therefore, most likely to have spiral progenitors that faded without experiencing major mergers) is higher in massive environments, such as clusters and big groups. Similarly, the fraction of galaxies that are more pressure supported (and therefore most likely to be the product of mergers) is higher in less massive environments, such as small groups or isolated galaxies.

In agreement with previous studies, we found that the fraction of galaxies formed through the "faded spiral" pathway is larger in the cluster than in the field. This is consistent with the idea that ram pressure stripping events contribute in clusters and not in the field. However, in the field, we have found that the "faded spiral" is still the dominant formation mechanism. This can be explained by secular evolution processes \citep{Kormendy13}, by assuming high efficiency of feedback processes in the field environment \citep{Li+18, Torrey+20}, or by a contamination of field spiral galaxies in our sample. The latter hypothesis is discussed in Appendix \ref{app:selection_biases}.

In our work, we extended the parameter space to infer the most probable formation pathway, rather than relying on kinematics only. On the other hand, we see in Fig. \ref{fig:clustering} that the kinematic parameters (and maybe mass) are those that show higher bimodality, enabling a better separation between the two classes of S0s (associated with the two formation pathways).
We also showed that the relation between formation pathways and the environment is also "local", and it depends on the density of galaxies in that environment. In the past this was probed by using  the properties of the stellar populations as indicators of formation scenario. \citet{McDermid+15} found a weak indication that older systems live in less dense environments, whereas \citet{Fraser+18} did not find significant correlations.

Our results are partially consistent with the simulation analysed by \citet{Deeley+21}. In their simulations, S0s are identified visually, by creating simulated images via a radiative transfer code and rescaling them to match the angular sizes in the SAMI survey. In their work, they found that the majority of the simulated S0s were produced via the "merger" scenario, whereas the majority of our observed S0s is formed via the "faded spiral" scenario. They also find that the merger scenario becomes more efficient with decreasing environment mass, in agreement with our results. 
The differences between our results and the simulations might be caused by the fact that i) our sample is not complete; ii) it is not possible to have a 1:1 association between galaxy properties and formation scenario; and iii) intrinsic differences between real galaxy images and those obtained from the simulations.

 Our finding that the fading spiral scenario is the most efficient in producing S0s might be at odd with previous observations.
\citet{Burstein+05} show that S0s galaxies, as a class, cannot be simply derived by removing gas from a spiral progenitor, because this would imply S0s to be at least 0.75 magnitudes fainter than Spirals, whereas they are as luminous than spirals (if not more luminous). But, when comparing S0s and spirals {\it of the same mass}, the picture described in \citet{Burstein+05} changes. Indeed, \citet{Williams+10} found that S0s in the local universe are systematically fainter by $0.53 \pm 0.15$ spiral galaxies of the same rotational velocity (which is a proxy for mass). However, always according to \citet{Williams+10}, the observed magnitude offset still does not support the scenario where high redshift spirals are the progenitors of present day S0s, because the time needed to fade by 0.5 mags would be $\sim 1.4$ Gyr at most (assuming constant star formation rate). This roughly corresponds to $z\sim 0.1$, but S0s are in place already at higher redshifts (e.g., \citealt{Fasano+00, Jaffe+14, denHeijer+15}). 

This means that, while the faded-spiral scenario could indeed be considered a viable mechanisms for the formation of S0s at low redshifts (on which our work is based), it has problems in explaining the formation of S0s at higher redshifts. One would require either a much more complex star formation or that the formation process is more complicated than a simple fading. The latter hypothesis is not surprising, because even after an S0 formed at high redshift from a spiral, it can experience further events such as rejuvenation or even mergers.  Therefore, the fading-spiral scenario for an S0 does not exclude mergers events during that galaxy's lifetime. It simply indicates that the removal or consumption of gas from a spiral progenitor are the processes that mostly contributed to the present day properties of that galaxy (at least to the properties presented in Table \ref{tab:sample}). Thus, it would be important to study how much the efficiency of the two formation scenarios change with redshift.

\section{Summary and conclusions}
\label{sec:conclusion}

In this paper we have inspected the properties and the environment of a large sample of lenticular galaxies. We divided them into two classes, on the basis of their physical properties. These two classes can be associated to the final products of two formation pathways: a "faded spiral scenario", that includes younger, less massive, less pressure supported, lower B/T, and lower S\`ersic indices galaxies (i.e. S0s of "Class A") and a "merger scenario", that includes older, more massive, more pressure supported, higher B/T, and higher S\`ersic index galaxies (i.e. S0s of "Class B").

These two classes are distributed differently in different environments. If we characterise the environment with the local galaxy density (local environment) and total mass of the host halo (global environment) we see that the "faded spiral" pathway is the most efficient channel in producing S0s, its efficiency increasing with the increase of the global environment mass or the local galaxy number density. The merger pathway is also a viable channel, and its efficiency becomes higher with decreasing local galaxy density or environment mass. The dependency on the formation mechanism is more evident when we consider the "local" environment (i.e. galaxy density) rather than the "global" environment (i.e., total mass of the host halo).

Even if we have considered only two formation channels in our work, the actual physical mechanisms that drives a given process might be different in different environments. For example, gas removal processes included in the "faded-spiral" scenario are caused by different mechanisms in clusters (e.g. ram pressure stripping experienced by the infalling galaxy) and in the field (e.g. AGN or SN feedback). Also, both processes (mergers and fading) can happen during a galaxy lifetime. A spiral disk can form at high redshift through merger and then passively fade into an S0s in isolation, or experience ram stripping when entering a cluster. At the same time, an S0 could be derived by a spiral progenitor and experience a series of mergers only recently. Our analysis indicates what was the path that had more impact in setting the properties we observe today (at least, for the properties considered in this study).

There is one  additional aspect that has to be considered when interpreting observational results. The signatures of the formation process, the properties and the morphology of a galaxy, or even the location of a S0 galaxy in a given environment change with time. However, our study and those from the literature are based on galaxy properties at the time of observation. This means that fraction of galaxies of a given class in a specific environment (and therefore the efficiency of a formation pathway in a specific environment) might have changed with redshift, as suggested in Section \ref{sec:comparison}. We therefore conclude that, while there is no well defined parameter threshold that uniquely associates a galaxy with a formation scenario, the environment plays a fundamental role in deciding the formation path of lenticular galaxies,  despite its signature in the properties of a galaxy might has become weaker as the galaxy evolves. A follow-up  with a large sample of S0s in different environments and at different redshifts could be very useful in understanding the redshift evolution of the efficiency of the formation processes of lenticular galaxies. 

\section*{Data Availability}
The data underlying this article were accessed from the MaNGA (https://www.sdss.org/dr13/manga/manga-data/) and SAMI (https://sami-survey.org/abdr) surveys. The derived data generated in this research
will be shared on reasonable request to the corresponding author.

\section*{Acknowledgments}
The authors thank Simon Deleey for useful discussion and his availability to share some of his results for cross-checking. 
E.J.J. acknowledges support from FONDECYT Iniciaci\'on en investigaci\'on Project 11200263.
YJ acknowledges finantial support from FONDECYT Iniciacion
2018 No. 11180558 and the ANID BASAL project FB210003

The SAMI Galaxy Survey is based on observations made at the Anglo-Australian Telescope. The Sydney-AAO Multi-object Integral field spectrograph (SAMI) was developed jointly by the University of Sydney and the Australian Astronomical Observatory. The SAMI input catalogue is based on data taken from the Sloan Digital Sky Survey, the GAMA Survey and the VST ATLAS Survey. The SAMI Galaxy Survey is funded by the Australian Research Council Centre of Excellence for All-sky Astrophysics (CAASTRO), through project number CE110001020, and other participating institutions. The SAMI Galaxy Survey website is http://sami-survey.org/.

Funding for the Sloan Digital Sky  Survey IV has been provided by the  Alfred P. Sloan Foundation, the U.S. Department of Energy Office of  Science, and the Participating  Institutions. 
SDSS-IV acknowledges support and resources from the Center for High Performance Computing  at the University of Utah. The SDSS website is www.sdss.org.
SDSS-IV is managed by the Astrophysical Research Consortium for the Participating Institutions 
of the SDSS Collaboration including the Brazilian Participation Group, the Carnegie Institution for Science, Carnegie Mellon University, Center for Astrophysics | Harvard \& Smithsonian, the Chilean Participation Group, the French Participation Group, Instituto de Astrof\'isica de 
Canarias, The Johns Hopkins University, Kavli Institute for the Physics and Mathematics of the Universe (IPMU) / University of Tokyo, the Korean Participation Group, 
Lawrence Berkeley National Laboratory, Leibniz Institut f\"ur Astrophysik 
Potsdam (AIP),  Max-Planck-Institut f\"ur Astronomie (MPIA Heidelberg), 
Max-Planck-Institut f\"ur Astrophysik (MPA Garching), Max-Planck-Institut f\"ur Extraterrestrische Physik (MPE), National Astronomical Observatories of China, New Mexico State University, 
New York University, University of Notre Dame, Observat\'ario Nacional / MCTI, The Ohio State University, Pennsylvania State University, Shanghai Astronomical Observatory, United 
Kingdom Participation Group, Universidad Nacional Aut\'onoma 
de M\'exico, University of Arizona, University of Colorado Boulder, 
University of Oxford, University of Portsmouth, University of Utah, 
University of Virginia, University of Washington, University of 
Wisconsin, Vanderbilt University, and Yale University.



\bibliographystyle{mnras}
\bibliography{S0III} 

\begin{thebibliography}{}
\makeatletter
\relax
\def\mn@urlcharsother{\let\do\@makeother \do\$\do\&\do\#\do\^\do\_\do\%\do\~}
\def\mn@doi{\begingroup\mn@urlcharsother \@ifnextchar [ {\mn@doi@}
  {\mn@doi@[]}}
\def\mn@doi@[#1]#2{\def\@tempa{#1}\ifx\@tempa\@empty \href
  {http://dx.doi.org/#2} {doi:#2}\else \href {http://dx.doi.org/#2} {#1}\fi
  \endgroup}
\def\mn@eprint#1#2{\mn@eprint@#1:#2::\@nil}
\def\mn@eprint@arXiv#1{\href {http://arxiv.org/abs/#1} {{\tt arXiv:#1}}}
\def\mn@eprint@dblp#1{\href {http://dblp.uni-trier.de/rec/bibtex/#1.xml}
  {dblp:#1}}
\def\mn@eprint@#1:#2:#3:#4\@nil{\def\@tempa {#1}\def\@tempb {#2}\def\@tempc
  {#3}\ifx \@tempc \@empty \let \@tempc \@tempb \let \@tempb \@tempa \fi \ifx
  \@tempb \@empty \def\@tempb {arXiv}\fi \@ifundefined
  {mn@eprint@\@tempb}{\@tempb:\@tempc}{\expandafter \expandafter \csname
  mn@eprint@\@tempb\endcsname \expandafter{\@tempc}}}

\bibitem[\protect\citeauthoryear{{Aguado} et~al.,}{{Aguado}
  et~al.}{2019}]{Aguado+19}
{Aguado} D.~S.,  et~al., 2019, \mn@doi [\apjs] {10.3847/1538-4365/aaf651},
  \href {https://ui.adsabs.harvard.edu/abs/2019ApJS..240...23A} {240, 23}

\bibitem[\protect\citeauthoryear{{Bekki} \& {Couch}}{{Bekki} \&
  {Couch}}{2011}]{Bekki+11}
{Bekki} K.,  {Couch} W.~J.,  2011, \mn@doi [\mnras]
  {10.1111/j.1365-2966.2011.18821.x}, \href
  {https://ui.adsabs.harvard.edu/abs/2011MNRAS.415.1783B} {415, 1783}

\bibitem[\protect\citeauthoryear{{Boquien}, {Burgarella}, {Roehlly}, {Buat},
  {Ciesla}, {Corre}, {Inoue}  \& {Salas}}{{Boquien} et~al.}{2019}]{Boquien+19}
{Boquien} M.,  {Burgarella} D.,  {Roehlly} Y.,  {Buat} V.,  {Ciesla} L.,
  {Corre} D.,  {Inoue} A.~K.,   {Salas} H.,  2019, \mn@doi [\aap]
  {10.1051/0004-6361/201834156}, \href
  {https://ui.adsabs.harvard.edu/abs/2019A&A...622A.103B} {622, A103}

\bibitem[\protect\citeauthoryear{{Brough} et~al.,}{{Brough}
  et~al.}{2013}]{Brough+13}
{Brough} S.,  et~al., 2013, \mn@doi [\mnras] {10.1093/mnras/stt1489}, \href
  {https://ui.adsabs.harvard.edu/abs/2013MNRAS.435.2903B} {435, 2903}

\bibitem[\protect\citeauthoryear{{Bryant} et~al.,}{{Bryant}
  et~al.}{2015}]{Bryant+15}
{Bryant} J.~J.,  et~al., 2015, \mn@doi [\mnras] {10.1093/mnras/stu2635}, \href
  {https://ui.adsabs.harvard.edu/abs/2015MNRAS.447.2857B} {447, 2857}

\bibitem[\protect\citeauthoryear{{Bundy} et~al.,}{{Bundy}
  et~al.}{2015}]{Bundy+15}
{Bundy} K.,  et~al., 2015, \mn@doi [ApJ] {10.1088/0004-637X/798/1/7}, \href
  {https://ui.adsabs.harvard.edu/abs/2015ApJ...798....7B} {798, 7}

\bibitem[\protect\citeauthoryear{{Burstein}, {Ho}, {Huchra}  \&
  {Macri}}{{Burstein} et~al.}{2005}]{Burstein+05}
{Burstein} D.,  {Ho} L.~C.,  {Huchra} J.~P.,   {Macri} L.~M.,  2005, \mn@doi
  [\apj] {10.1086/427408}, \href
  {https://ui.adsabs.harvard.edu/abs/2005ApJ...621..246B} {621, 246}

\bibitem[\protect\citeauthoryear{{Cappellari}}{{Cappellari}}{2017}]{Cappellari+17}
{Cappellari} M.,  2017, \mn@doi [\mnras] {10.1093/mnras/stw3020}, \href
  {https://ui.adsabs.harvard.edu/abs/2017MNRAS.466..798C} {466, 798}

\bibitem[\protect\citeauthoryear{{Cappellari} \& {Emsellem}}{{Cappellari} \&
  {Emsellem}}{2004}]{Cappellari+04}
{Cappellari} M.,  {Emsellem} E.,  2004, \mn@doi [\pasp] {10.1086/381875}, \href
  {https://ui.adsabs.harvard.edu/abs/2004PASP..116..138C} {116, 138}

\bibitem[\protect\citeauthoryear{{Cappellari} et~al.,}{{Cappellari}
  et~al.}{2011}]{Cappellari+11}
{Cappellari} M.,  et~al., 2011, \mn@doi [\mnras]
  {10.1111/j.1365-2966.2011.18600.x}, \href
  {https://ui.adsabs.harvard.edu/abs/2011MNRAS.416.1680C} {416, 1680}

\bibitem[\protect\citeauthoryear{{Catinella} et~al.,}{{Catinella}
  et~al.}{2012}]{Catinella+12}
{Catinella} B.,  et~al., 2012, \mn@doi [\mnras]
  {10.1111/j.1365-2966.2011.20012.x}, \href
  {https://ui.adsabs.harvard.edu/abs/2012MNRAS.420.1959C} {420, 1959}

\bibitem[\protect\citeauthoryear{{Chilingarian}, {Di Matteo}, {Combes},
  {Melchior}  \& {Semelin}}{{Chilingarian} et~al.}{2010}]{Chilingarian+10}
{Chilingarian} I.~V.,  {Di Matteo} P.,  {Combes} F.,  {Melchior} A.~L.,
  {Semelin} B.,  2010, \mn@doi [\aap] {10.1051/0004-6361/200912938}, \href
  {https://ui.adsabs.harvard.edu/abs/2010A&A...518A..61C} {518, A61}

\bibitem[\protect\citeauthoryear{{Coccato} et~al.,}{{Coccato}
  et~al.}{2020}]{Coccato+20}
{Coccato} L.,  et~al., 2020, \mn@doi [\mnras] {10.1093/mnras/stz3592}, \href
  {https://ui.adsabs.harvard.edu/abs/2020MNRAS.492.2955C} {492, 2955}

\bibitem[\protect\citeauthoryear{{Cortese} et~al.,}{{Cortese}
  et~al.}{2014}]{Cortese+14}
{Cortese} L.,  et~al., 2014, \mn@doi [\apjl] {10.1088/2041-8205/795/2/L37},
  \href {https://ui.adsabs.harvard.edu/abs/2014ApJ...795L..37C} {795, L37}

\bibitem[\protect\citeauthoryear{{Cortese} et~al.,}{{Cortese}
  et~al.}{2016}]{Cortese+16}
{Cortese} L.,  et~al., 2016, \mn@doi [\mnras] {10.1093/mnras/stw1891}, \href
  {https://ui.adsabs.harvard.edu/abs/2016MNRAS.463..170C} {463, 170}

\bibitem[\protect\citeauthoryear{{Croom} et~al.,}{{Croom}
  et~al.}{2021}]{Croom+21}
{Croom} S.~M.,  et~al., 2021, \mn@doi [\mnras] {10.1093/mnras/stab229}, \href
  {https://ui.adsabs.harvard.edu/abs/2021MNRAS.505..991C} {505, 991}

\bibitem[\protect\citeauthoryear{{Dav{\'e}}, {Angl{\'e}s-Alc{\'a}zar},
  {Narayanan}, {Li}, {Rafieferantsoa}  \& {Appleby}}{{Dav{\'e}}
  et~al.}{2019}]{Dave+19}
{Dav{\'e}} R.,  {Angl{\'e}s-Alc{\'a}zar} D.,  {Narayanan} D.,  {Li} Q.,
  {Rafieferantsoa} M.~H.,   {Appleby} S.,  2019, \mn@doi [\mnras]
  {10.1093/mnras/stz937}, \href
  {https://ui.adsabs.harvard.edu/abs/2019MNRAS.486.2827D} {486, 2827}

\bibitem[\protect\citeauthoryear{{Deeley} et~al.,}{{Deeley}
  et~al.}{2020}]{Deeley+20}
{Deeley} S.,  et~al., 2020, \mn@doi [\mnras] {10.1093/mnras/staa2417}, \href
  {https://ui.adsabs.harvard.edu/abs/2020MNRAS.498.2372D} {498, 2372}

\bibitem[\protect\citeauthoryear{{Deeley}, {Drinkwater}, {Sweet}, {Bekki},
  {Couch}, {Forbes}  \& {Dolfi}}{{Deeley} et~al.}{2021}]{Deeley+21}
{Deeley} S.,  {Drinkwater} M.~J.,  {Sweet} S.~M.,  {Bekki} K.,  {Couch} W.~J.,
  {Forbes} D.~A.,   {Dolfi} A.,  2021, \mn@doi [\mnras]
  {10.1093/mnras/stab2007}, \href
  {https://ui.adsabs.harvard.edu/abs/2021MNRAS.508..895D} {508, 895}

\bibitem[\protect\citeauthoryear{{Dey} et~al.,}{{Dey} et~al.}{2019}]{Dey+19}
{Dey} A.,  et~al., 2019, \mn@doi [\aj] {10.3847/1538-3881/ab089d}, \href
  {https://ui.adsabs.harvard.edu/abs/2019AJ....157..168D} {157, 168}

\bibitem[\protect\citeauthoryear{{Dressler}}{{Dressler}}{1980}]{Dressler80}
{Dressler} A.,  1980, \mn@doi [\apj] {10.1086/157753}, \href
  {https://ui.adsabs.harvard.edu/abs/1980ApJ...236..351D} {236, 351}

\bibitem[\protect\citeauthoryear{{Eliche-Moral}, {Rodr{\'\i}guez-P{\'e}rez},
  {Borlaff}, {Querejeta}  \& {Tapia}}{{Eliche-Moral}
  et~al.}{2018}]{Eliche-Moral+2018}
{Eliche-Moral} M.~C.,  {Rodr{\'\i}guez-P{\'e}rez} C.,  {Borlaff} A.,
  {Querejeta} M.,   {Tapia} T.,  2018, \mn@doi [\aap]
  {10.1051/0004-6361/201832911}, \href
  {https://ui.adsabs.harvard.edu/abs/2018A&A...617A.113E} {617, A113}

\bibitem[\protect\citeauthoryear{{Emsellem} et~al.,}{{Emsellem}
  et~al.}{2007}]{Emsellem+07}
{Emsellem} E.,  et~al., 2007, \mn@doi [\mnras]
  {10.1111/j.1365-2966.2007.11752.x}, \href
  {https://ui.adsabs.harvard.edu/abs/2007MNRAS.379..401E} {379, 401}

\bibitem[\protect\citeauthoryear{{Emsellem} et~al.,}{{Emsellem}
  et~al.}{2011}]{Emsellem+11}
{Emsellem} E.,  et~al., 2011, \mn@doi [\mnras]
  {10.1111/j.1365-2966.2011.18496.x}, \href
  {https://ui.adsabs.harvard.edu/abs/2011MNRAS.414..888E} {414, 888}

\bibitem[\protect\citeauthoryear{{Fasano}, {Poggianti}, {Couch}, {Bettoni},
  {Kj{\ae}rgaard}  \& {Moles}}{{Fasano} et~al.}{2000}]{Fasano+00}
{Fasano} G.,  {Poggianti} B.~M.,  {Couch} W.~J.,  {Bettoni} D.,
  {Kj{\ae}rgaard} P.,   {Moles} M.,  2000, \mn@doi [\apj] {10.1086/317047},
  \href {https://ui.adsabs.harvard.edu/abs/2000ApJ...542..673F} {542, 673}

\bibitem[\protect\citeauthoryear{{Fraser-McKelvie}, {Arag{\'o}n-Salamanca},
  {Merrifield}, {Tabor}, {Bernardi}, {Drory}, {Parikh}  \&
  {Argudo-Fern{\'a}ndez}}{{Fraser-McKelvie} et~al.}{2018}]{Fraser+18}
{Fraser-McKelvie} A.,  {Arag{\'o}n-Salamanca} A.,  {Merrifield} M.,  {Tabor}
  M.,  {Bernardi} M.,  {Drory} N.,  {Parikh} T.,   {Argudo-Fern{\'a}ndez} M.,
  2018, \mn@doi [\mnras] {10.1093/mnras/sty2563}, \href
  {https://ui.adsabs.harvard.edu/abs/2018MNRAS.481.5580F} {481, 5580}

\bibitem[\protect\citeauthoryear{{Fraser-McKelvie} et~al.,}{{Fraser-McKelvie}
  et~al.}{2021}]{Fraser+21}
{Fraser-McKelvie} A.,  et~al., 2021, \mn@doi [\mnras] {10.1093/mnras/stab573},
  \href {https://ui.adsabs.harvard.edu/abs/2021MNRAS.503.4992F} {503, 4992}

\bibitem[\protect\citeauthoryear{{Governato}, {Willman}, {Mayer}, {Brooks},
  {Stinson}, {Valenzuela}, {Wadsley}  \& {Quinn}}{{Governato}
  et~al.}{2007}]{Governato+07}
{Governato} F.,  {Willman} B.,  {Mayer} L.,  {Brooks} A.,  {Stinson} G.,
  {Valenzuela} O.,  {Wadsley} J.,   {Quinn} T.,  2007, \mn@doi [\mnras]
  {10.1111/j.1365-2966.2006.11266.x}, \href
  {https://ui.adsabs.harvard.edu/abs/2007MNRAS.374.1479G} {374, 1479}

\bibitem[\protect\citeauthoryear{{Harborne}, {van de Sande}, {Cortese},
  {Power}, {Robotham}, {Lagos}  \& {Croom}}{{Harborne}
  et~al.}{2020}]{Harborne+20}
{Harborne} K.~E.,  {van de Sande} J.,  {Cortese} L.,  {Power} C.,  {Robotham}
  A.~S.~G.,  {Lagos} C.~D.~P.,   {Croom} S.,  2020, \mn@doi [\mnras]
  {10.1093/mnras/staa1847}, \href
  {https://ui.adsabs.harvard.edu/abs/2020MNRAS.497.2018H} {497, 2018}

\bibitem[\protect\citeauthoryear{{Jackson}, {Kaviraj}, {Martin}, {Devriendt},
  {Noakes-Kettel}, {Silk}, {Ogle}  \& {Dubois}}{{Jackson}
  et~al.}{2022}]{Jackson+22}
{Jackson} R.~A.,  {Kaviraj} S.,  {Martin} G.,  {Devriendt} J.~E.~G.,
  {Noakes-Kettel} E.~A.,  {Silk} J.,  {Ogle} P.,   {Dubois} Y.,  2022, \mn@doi
  [\mnras] {10.1093/mnras/stac058}, \href
  {https://ui.adsabs.harvard.edu/abs/2022MNRAS.511..607J} {511, 607}

\bibitem[\protect\citeauthoryear{{Jaff{\'e}} et~al.,}{{Jaff{\'e}}
  et~al.}{2014}]{Jaffe+14}
{Jaff{\'e}} Y.~L.,  et~al., 2014, \mn@doi [\mnras] {10.1093/mnras/stu507},
  \href {https://ui.adsabs.harvard.edu/abs/2014MNRAS.440.3491J} {440, 3491}

\bibitem[\protect\citeauthoryear{{Johnston} et~al.,}{{Johnston}
  et~al.}{2021}]{Johnston+21}
{Johnston} E.~J.,  et~al., 2021, \mn@doi [\mnras] {10.1093/mnras/staa2838},
  \href {https://ui.adsabs.harvard.edu/abs/2021MNRAS.500.4193J} {500, 4193}

\bibitem[\protect\citeauthoryear{{Karachentseva}, {Mitronova}, {Melnyk}  \&
  {Karachentsev}}{{Karachentseva} et~al.}{2010}]{Karachentseva+10}
{Karachentseva} V.~E.,  {Mitronova} S.~N.,  {Melnyk} O.~V.,   {Karachentsev}
  I.~D.,  2010, \mn@doi [Astrophysical Bulletin] {10.1134/S1990341310010013},
  \href {https://ui.adsabs.harvard.edu/abs/2010AstBu..65....1K} {65, 1}

\bibitem[\protect\citeauthoryear{{Kormendy}}{{Kormendy}}{2013}]{Kormendy13}
{Kormendy} J.,  2013, in {Falc{\'o}n-Barroso} J.,  {Knapen} J.~H.,  eds, ,
  Secular Evolution of Galaxies.
p.~1

\bibitem[\protect\citeauthoryear{{Krajnovi{\'c}} et~al.,}{{Krajnovi{\'c}}
  et~al.}{2011}]{Krajnovic+11}
{Krajnovi{\'c}} D.,  et~al., 2011, \mn@doi [\mnras]
  {10.1111/j.1365-2966.2011.18560.x}, \href
  {https://ui.adsabs.harvard.edu/abs/2011MNRAS.414.2923K} {414, 2923}

\bibitem[\protect\citeauthoryear{{Laurikainen}, {Salo}, {Buta}, {Knapen}  \&
  {Comer{\'o}n}}{{Laurikainen} et~al.}{2010}]{Laurikainen+10}
{Laurikainen} E.,  {Salo} H.,  {Buta} R.,  {Knapen} J.~H.,   {Comer{\'o}n} S.,
  2010, \mn@doi [\mnras] {10.1111/j.1365-2966.2010.16521.x}, \href
  {https://ui.adsabs.harvard.edu/abs/2010MNRAS.405.1089L} {405, 1089}

\bibitem[\protect\citeauthoryear{{Li} et~al.,}{{Li} et~al.}{2018}]{Li+18}
{Li} Y.-P.,  et~al., 2018, \mn@doi [\apj] {10.3847/1538-4357/aade8b}, \href
  {https://ui.adsabs.harvard.edu/abs/2018ApJ...866...70L} {866, 70}

\bibitem[\protect\citeauthoryear{{McDermid} et~al.,}{{McDermid}
  et~al.}{2015}]{McDermid+15}
{McDermid} R.~M.,  et~al., 2015, \mn@doi [\mnras] {10.1093/mnras/stv105}, \href
  {https://ui.adsabs.harvard.edu/abs/2015MNRAS.448.3484M} {448, 3484}

\bibitem[\protect\citeauthoryear{{Mu{\~n}oz-Mateos}, {Gil de Paz}, {Boissier},
  {Zamorano}, {Jarrett}, {Gallego}  \& {Madore}}{{Mu{\~n}oz-Mateos}
  et~al.}{2007}]{munoz+07}
{Mu{\~n}oz-Mateos} J.~C.,  {Gil de Paz} A.,  {Boissier} S.,  {Zamorano} J.,
  {Jarrett} T.,  {Gallego} J.,   {Madore} B.~F.,  2007, \mn@doi [\apj]
  {10.1086/511812}, \href
  {https://ui.adsabs.harvard.edu/abs/2007ApJ...658.1006M} {658, 1006}

\bibitem[\protect\citeauthoryear{{Navarro} \& {White}}{{Navarro} \&
  {White}}{1994}]{Navarro+94}
{Navarro} J.~F.,  {White} S. D.~M.,  1994, \mn@doi [\mnras]
  {10.1093/mnras/267.2.401}, \href
  {https://ui.adsabs.harvard.edu/abs/1994MNRAS.267..401N} {267, 401}

\bibitem[\protect\citeauthoryear{{Noll}, {Burgarella}, {Giovannoli}, {Buat},
  {Marcillac}  \& {Mu{\~n}oz-Mateos}}{{Noll} et~al.}{2009}]{Noll+09}
{Noll} S.,  {Burgarella} D.,  {Giovannoli} E.,  {Buat} V.,  {Marcillac} D.,
  {Mu{\~n}oz-Mateos} J.~C.,  2009, \mn@doi [\aap]
  {10.1051/0004-6361/200912497}, \href
  {https://ui.adsabs.harvard.edu/abs/2009A&A...507.1793N} {507, 1793}

\bibitem[\protect\citeauthoryear{{Rathore}, {Kumar}, {Mishra}, {Wadadekar}  \&
  {Bait}}{{Rathore} et~al.}{2022}]{Rathore+22}
{Rathore} H.,  {Kumar} K.,  {Mishra} P.~K.,  {Wadadekar} Y.,   {Bait} O.,
  2022, \mn@doi [\mnras] {10.1093/mnras/stac871}, \href
  {https://ui.adsabs.harvard.edu/abs/2022MNRAS.tmp..868R} {}

\bibitem[\protect\citeauthoryear{{Rizzo}, {Fraternali}  \& {Iorio}}{{Rizzo}
  et~al.}{2018}]{Rizzo+18}
{Rizzo} F.,  {Fraternali} F.,   {Iorio} G.,  2018, \mn@doi [\mnras]
  {10.1093/mnras/sty347}, \href
  {https://ui.adsabs.harvard.edu/abs/2018MNRAS.476.2137R} {476, 2137}

\bibitem[\protect\citeauthoryear{{Robotham} et~al.,}{{Robotham}
  et~al.}{2011}]{Robotham+11}
{Robotham} A.~S.~G.,  et~al., 2011, \mn@doi [\mnras]
  {10.1111/j.1365-2966.2011.19217.x}, \href
  {https://ui.adsabs.harvard.edu/abs/2011MNRAS.416.2640R} {416, 2640}

\bibitem[\protect\citeauthoryear{{Salim} et~al.,}{{Salim}
  et~al.}{2016}]{Salim+16}
{Salim} S.,  et~al., 2016, \mn@doi [\apjs] {10.3847/0067-0049/227/1/2}, \href
  {https://ui.adsabs.harvard.edu/abs/2016ApJS..227....2S} {227, 2}

\bibitem[\protect\citeauthoryear{{Salim}, {Boquien}  \& {Lee}}{{Salim}
  et~al.}{2018}]{Salim+18}
{Salim} S.,  {Boquien} M.,   {Lee} J.~C.,  2018, \mn@doi [\apj]
  {10.3847/1538-4357/aabf3c}, \href
  {https://ui.adsabs.harvard.edu/abs/2018ApJ...859...11S} {859, 11}

\bibitem[\protect\citeauthoryear{{Simard}, {Mendel}, {Patton}, {Ellison}  \&
  {McConnachie}}{{Simard} et~al.}{2011}]{Simard+11}
{Simard} L.,  {Mendel} J.~T.,  {Patton} D.~R.,  {Ellison} S.~L.,
  {McConnachie} A.~W.,  2011, \mn@doi [\apjs] {10.1088/0067-0049/196/1/11},
  \href {https://ui.adsabs.harvard.edu/abs/2011ApJS..196...11S} {196, 11}

\bibitem[\protect\citeauthoryear{{Tapia}, {Eliche-Moral}, {Aceves},
  {Rodr{\'\i}guez-P{\'e}rez}, {Borlaff}  \& {Querejeta}}{{Tapia}
  et~al.}{2017}]{Tapia+17}
{Tapia} T.,  {Eliche-Moral} M.~C.,  {Aceves} H.,  {Rodr{\'\i}guez-P{\'e}rez}
  C.,  {Borlaff} A.,   {Querejeta} M.,  2017, \mn@doi [\aap]
  {10.1051/0004-6361/201628821}, \href
  {https://ui.adsabs.harvard.edu/abs/2017A&A...604A.105T} {604, A105}

\bibitem[\protect\citeauthoryear{{Tempel}, {Tago}  \& {Liivam{\"a}gi}}{{Tempel}
  et~al.}{2012}]{Tempel+12}
{Tempel} E.,  {Tago} E.,   {Liivam{\"a}gi} L.~J.,  2012, \mn@doi [\aap]
  {10.1051/0004-6361/201118687}, \href
  {https://ui.adsabs.harvard.edu/abs/2012A&A...540A.106T} {540, A106}

\bibitem[\protect\citeauthoryear{{Torrey} et~al.,}{{Torrey}
  et~al.}{2020}]{Torrey+20}
{Torrey} P.,  et~al., 2020, \mn@doi [\mnras] {10.1093/mnras/staa2222}, \href
  {https://ui.adsabs.harvard.edu/abs/2020MNRAS.497.5292T} {497, 5292}

\bibitem[\protect\citeauthoryear{{Vazdekis}, {S{\'a}nchez-Bl{\'a}zquez},
  {Falc{\'o}n-Barroso}, {Cenarro}, {Beasley}, {Cardiel}, {Gorgas}  \&
  {Peletier}}{{Vazdekis} et~al.}{2010}]{Vazdekis+10}
{Vazdekis} A.,  {S{\'a}nchez-Bl{\'a}zquez} P.,  {Falc{\'o}n-Barroso} J.,
  {Cenarro} A.~J.,  {Beasley} M.~A.,  {Cardiel} N.,  {Gorgas} J.,   {Peletier}
  R.~F.,  2010, \mn@doi [\mnras] {10.1111/j.1365-2966.2010.16407.x}, \href
  {https://ui.adsabs.harvard.edu/abs/2010MNRAS.404.1639V} {404, 1639}

\bibitem[\protect\citeauthoryear{{V{\'a}zquez-Mata} et~al.,}{{V{\'a}zquez-Mata}
  et~al.}{2022}]{Vazquez-Mata+22}
{V{\'a}zquez-Mata} J.~A.,  et~al., 2022, \mn@doi [\mnras]
  {10.1093/mnras/stac635}, \href
  {https://ui.adsabs.harvard.edu/abs/2022MNRAS.512.2222V} {512, 2222}

\bibitem[\protect\citeauthoryear{{Williams}, {Bureau}  \&
  {Cappellari}}{{Williams} et~al.}{2010}]{Williams+10}
{Williams} M.~J.,  {Bureau} M.,   {Cappellari} M.,  2010, \mn@doi [\mnras]
  {10.1111/j.1365-2966.2010.17406.x}, \href
  {https://ui.adsabs.harvard.edu/abs/2010MNRAS.409.1330W} {409, 1330}

\bibitem[\protect\citeauthoryear{{Yoon}, {Park}, {Chung}  \& {Lane}}{{Yoon}
  et~al.}{2021}]{Yoon+21}
{Yoon} Y.,  {Park} C.,  {Chung} H.,   {Lane} R.~R.,  2021, arXiv e-prints,
  \href {https://ui.adsabs.harvard.edu/abs/2021arXiv211213703Y} {p.
  arXiv:2112.13703}

\bibitem[\protect\citeauthoryear{{den Heijer} et~al.,}{{den Heijer}
  et~al.}{2015}]{denHeijer+15}
{den Heijer} M.,  et~al., 2015, \mn@doi [\aap] {10.1051/0004-6361/201526879},
  \href {https://ui.adsabs.harvard.edu/abs/2015A&A...581A..98D} {581, A98}

\bibitem[\protect\citeauthoryear{{van den Bergh}}{{van den
  Bergh}}{2009}]{vanDenBergh09}
{van den Bergh} S.,  2009, \mn@doi [\apj] {10.1088/0004-637X/702/2/1502}, \href
  {https://ui.adsabs.harvard.edu/abs/2009ApJ...702.1502V} {702, 1502}

\makeatother
\end{thebibliography}

%
\appendix
\section{Selection biases: contamination from ellipticals and spirals}
\label{app:selection_biases}

In this Section we account for sample selection biases caused by having used catalogues based on visual morphology classification to identify S0s galaxies. Indeed, the two main contamination sources are flattened ellipticals (their elongated structure might mimic the presence of a disk) and spirals with faint arms (poor spatial resolution might not reveal them). In order to minimise the effect of the contamination, we tested our results by repeating the analysis on a sub-sample of galaxies obtained by applying two additional constraints to the sample selection. First, we requested that the galaxy is a fast rotator by imposing \lambdar$/\sqrt{e} > 0.3$ \citep{Emsellem+07}, where $e$ is the galaxy ellipticity. This minimised the presence of purely elliptical systems while leaving galaxies with a disk-like kinematic component \citep{Cappellari+11, Krajnovic+11} in the sample. Secondly, we imposed a limit to the specific star formation rate, being  $\log_{10}{\rm sSFR/[yr^{-1}]} \leq -10.5$. This value is twice the value needed to a galaxy to double its stellar mass in a Hubble time \citep{Dave+19} and it is about $1\sigma$ below the mean specific star formation rate measured in local spiral galaxies \citep{munoz+07}. This minimised the presence of star-forming spiral galaxies in our sample, but could remove also some star forming S0s, whose recent star formation is believed to be triggered by minor mergers \citep{Rathore+22}.

\begin{figure*}
\includegraphics[width=17cm]{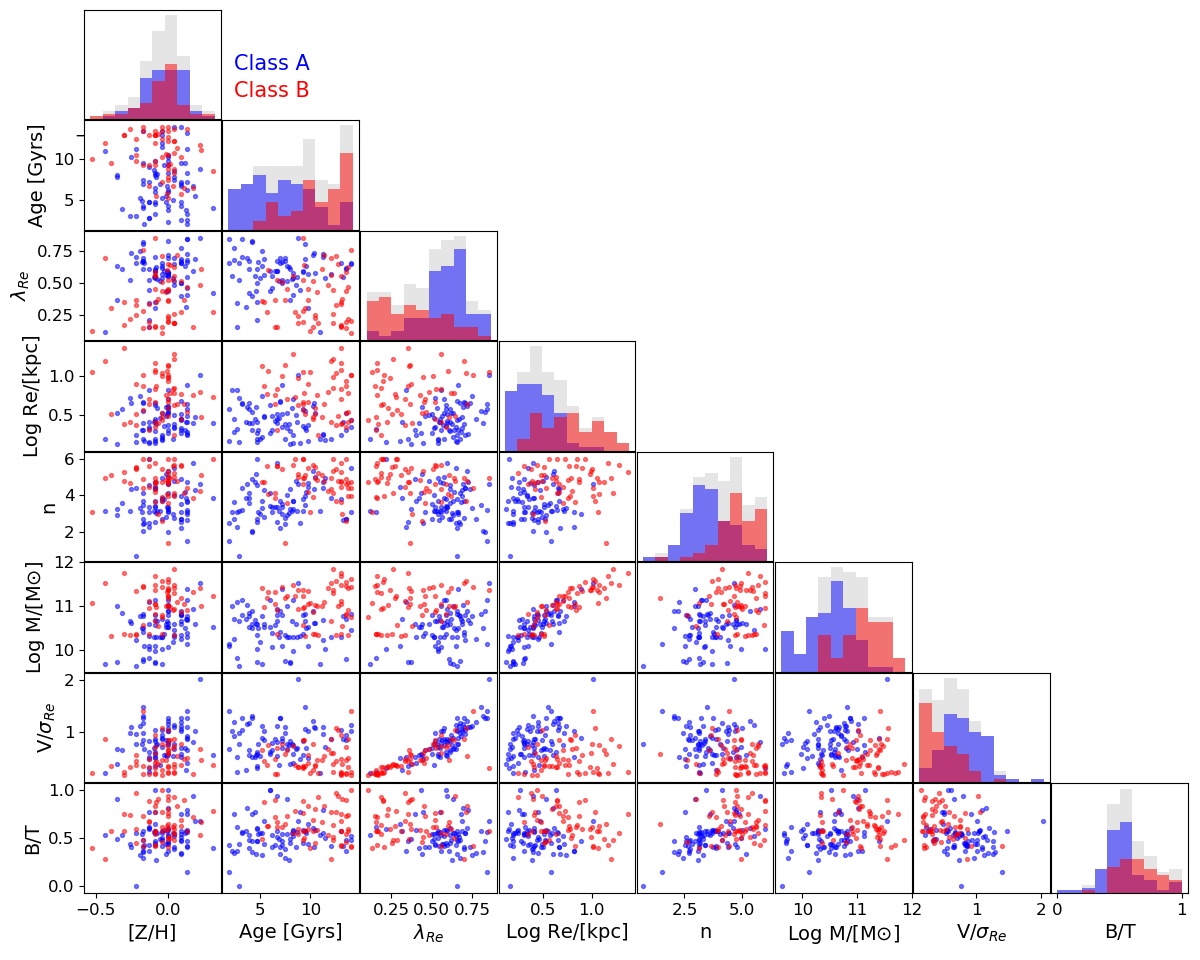}
\caption{As Fig. \ref{fig:clustering}, but considering only the 139 S0s that match the additional constrains on \lambdar\ and specific star formation rate (see text for details).}
\label{fig:clustering_app}
\end{figure*}

\begin{figure}
\psfig{file=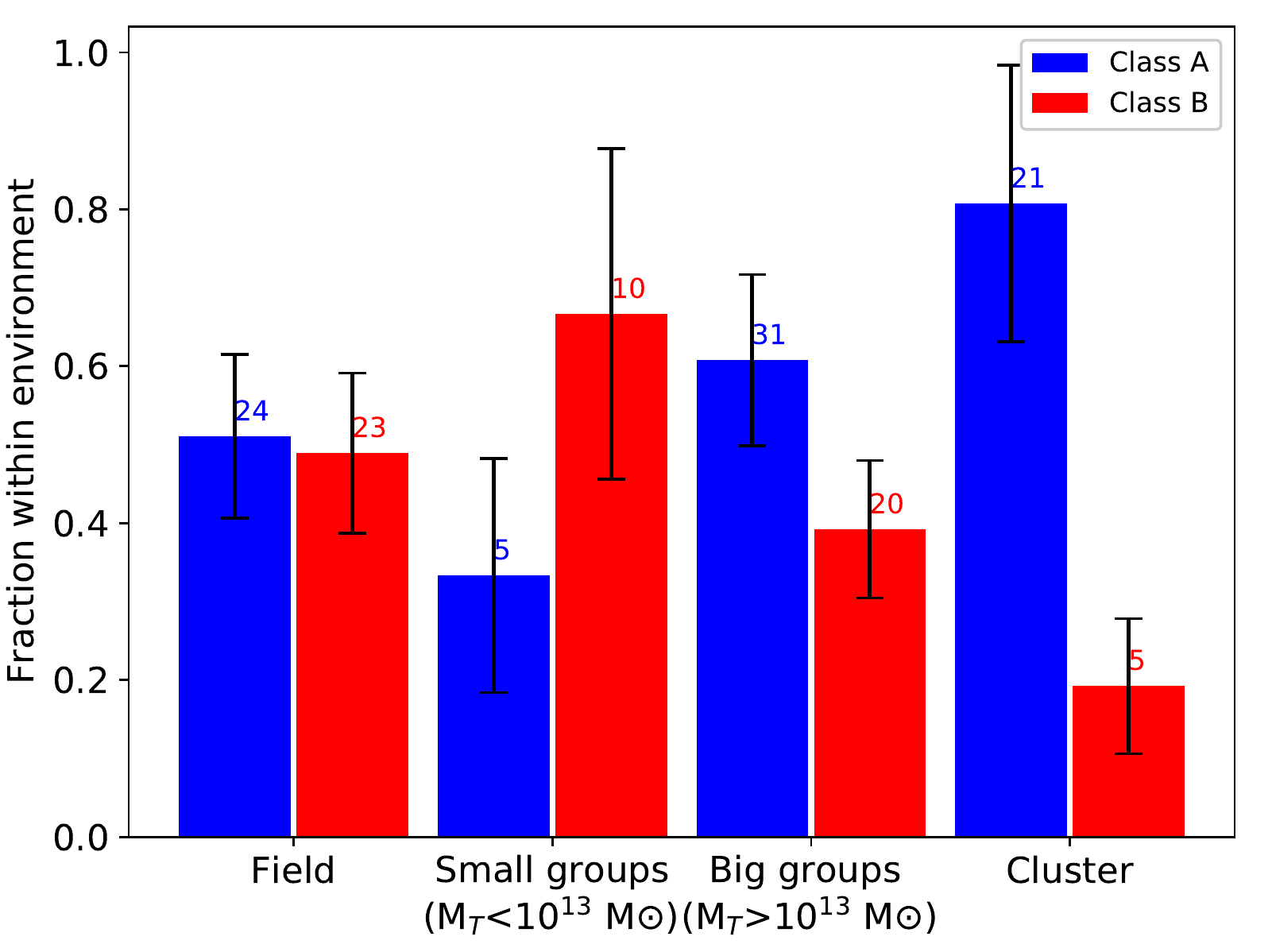,clip=,width=9cm}
\caption{As Fig. \ref{fig:env_mass}, but considering only the 139 S0s  that match the additional constrains on \lambdar\ and specific star formation rate (see text for details).}
\label{fig:env_mass_app}
\end{figure}

\begin{figure}
\hspace{-.5cm}
\psfig{file=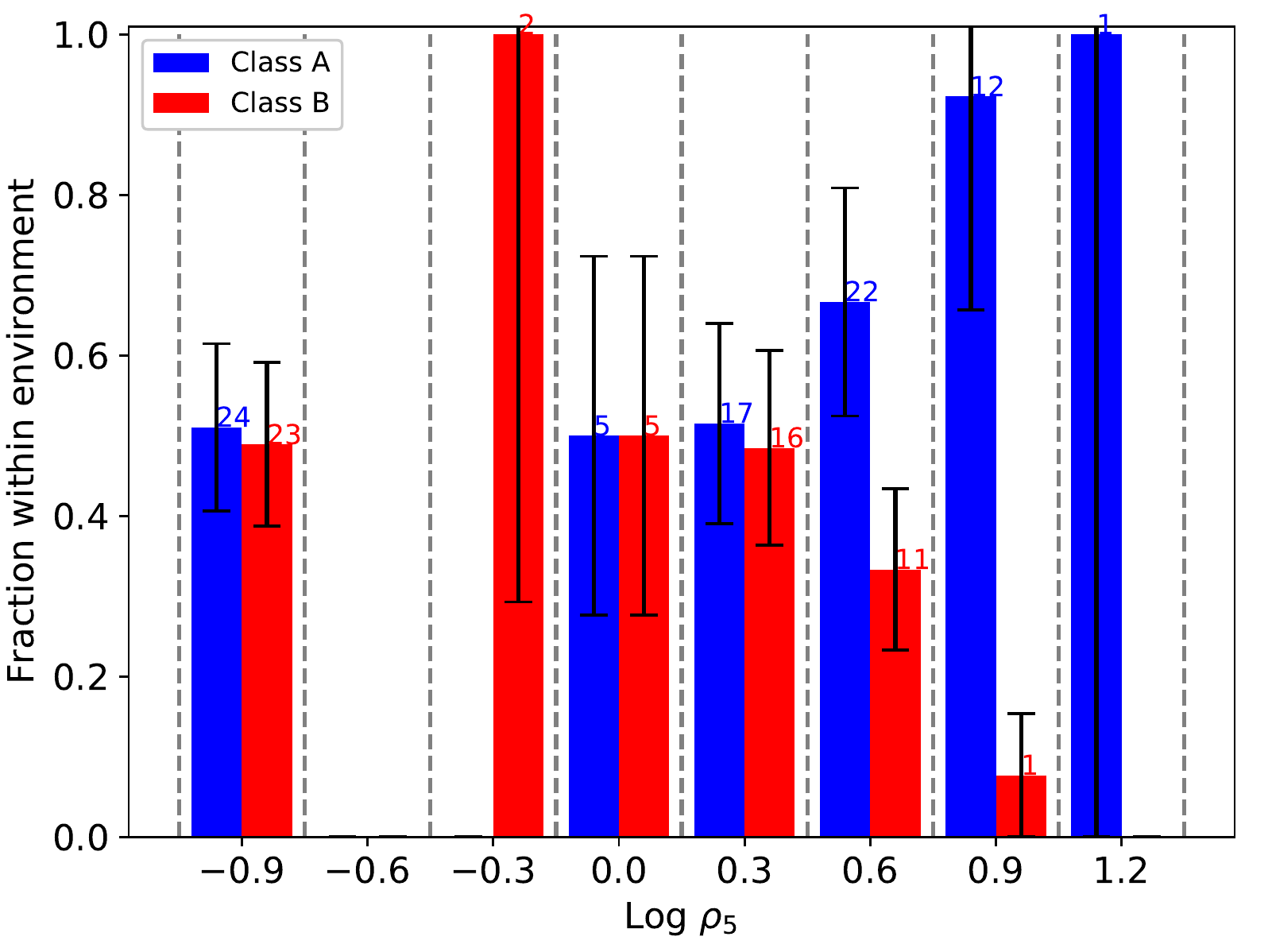,clip=,width=9cm}
\caption{As Fig. \ref{fig:density1}, but considering only the 139 S0s  that match the additional constrains on \lambdar\ and specific star formation rate (see text for details).}
\label{fig:density1_app}
\end{figure}

After applying these new selection criteria to the sample, we were left with a sub-sample of 139 S0s. We repeated our analysis and showed that most of our results still hold, despite the uncertainties begin higher. 
Figure \ref{fig:clustering_app} compares the physical properties considered in Section \ref{sec:clustering} to each other. Consistently with the previous results, we still find the presence of two populations of S0s. The first (class A) has properties consistent with S0s formed via the "faded-spiral" pathway. The second (class B) has properties consistent with S0s formed via the "merger" scenario.
Figures \ref{fig:env_mass_app} and \ref{fig:density1_app} show the distribution of class A and B S0s with the environment, where it is still possible to see the trend with environment properties as described in Section \ref{sec:formation_and_environment}. In contrast with the results of Section \ref{sec:formation_and_environment}, the "faded spiral" scenario does not dominate the field environment. This might indeed indicate a contamination of spiral galaxies in our selected sample, but given the high uncertainties, it is not possible to establish which process is more efficient in the field. 

%


\bsp	
\label{lastpage}
\end{document}